\documentclass[sigconf]{acmart}

\AtBeginDocument{%
  }

\copyrightyear{2022}
\acmYear{2022}
\setcopyright{acmlicensed}\acmConference[CIKM '22]{Proceedings of the 31st ACM International Conference on Information and Knowledge Management}{October 17--21, 2022}{Atlanta, GA, USA}
\acmBooktitle{Proceedings of the 31st ACM International Conference on Information and Knowledge Management (CIKM '22), October 17--21, 2022, Atlanta, GA, USA}
\acmPrice{15.00}
\acmDOI{10.1145/3511808.3557110}
\acmISBN{978-1-4503-9236-5/22/10}

\usepackage{multirow}
\usepackage{bm}
\usepackage{amsmath}
\usepackage{algorithmic}
\usepackage[linesnumbered,ruled,vlined]{algorithm2e}
\SetKwInput{KwInput}{Input}                
\SetKwInput{KwOutput}{Output}              
\SetKwInput{KwParameter}{Parameter}        
\usepackage{listings}
\usepackage{subfigure}

\newcommand{\specialcell}[2][c]{\begin{tabular}[#1]{@{}c@{}}#2\end{tabular}}

\DeclareMathOperator{\Tr}{Tr}
\DeclareMathOperator{\Cov}{Cov}

\begin{document}

\title[Simulation-Informed Revenue Extrapolation with Confidence Estimate \\ for Scaleup Companies Using Scarce Time-Series Data]{Simulation-Informed Revenue Extrapolation with Confidence Estimate for Scaleup Companies Using Scarce Time-Series Data}

\author{Lele Cao}
\authornote{Send correspondence to Lele Cao. The source code and anonymized benchmarking datasets are available at {\color{blue}\url{https://github.com/EQTPartners/sire}}}
\orcid{0000-0002-5680-9031}
\affiliation{%
  \institution{Motherbrain, EQT}
  \state{Stockholm}
  \country{Sweden}
  \postcode{11153}
}
\email{lele.cao@eqtpartners.com}

\author{Sonja Horn}
\orcid{0000-0003-3061-6845}
\affiliation{%
  \institution{Motherbrain, EQT}
  \state{Stockholm}
  \country{Sweden}
  \postcode{11153}
}
\email{sonja.horn@eqtpartners.com}

\author{Vilhelm von Ehrenheim}
\orcid{0000-0002-4210-4989}
\affiliation{%
  \institution{Motherbrain, EQT}
  \state{Stockholm}
  \country{Sweden}
  \postcode{11153}
}
\email{vilhelm.vonehrenheim@eqtpartners.com}

\author{Richard Anselmo Stahl}
\orcid{0000-0001-6008-8612}
\affiliation{%
 \institution{Motherbrain, EQT}
  \state{Stockholm}
  \country{Sweden}
  \postcode{11153}
}
\email{richard.stahl@eqtpartners.com}

\author{Henrik Landgren}
\orcid{0000-0002-2263-056X}
\affiliation{%
  \institution{Ark Kapital, Stockholm, Sweden}
  \country{}
  \postcode{11430}
}
\email{henrik@arkkapital.com}
\affiliation{%
  \institution{Motherbrain, EQT}
  \state{Stockholm}
  \country{Sweden}
  \postcode{11153}
}

\renewcommand{\shortauthors}{Lele Cao et al.}

\begin{abstract}
  Investment professionals rely on extrapolating company revenue into the future (i.e. revenue forecast) to approximate the valuation of scaleups (private companies in a high-growth stage) and inform their investment decision. This task is manual and empirical, leaving the forecast quality heavily dependent on the investment professionals' experiences and insights. Furthermore, financial data on scaleups is typically proprietary, costly and scarce, ruling out the wide adoption of data-driven approaches. To this end, we propose a simulation-informed revenue extrapolation (SiRE) algorithm that generates fine-grained long-term revenue predictions on small datasets and short time-series. SiRE models the revenue dynamics as a linear dynamical system (LDS), which is solved using the EM algorithm. The main innovation lies in how the noisy revenue measurements are obtained during training and inferencing. SiRE works for scaleups that operate in various sectors and provides confidence estimates. The quantitative experiments on two practical tasks show that SiRE significantly surpasses the baseline methods by a large margin. We also observe high performance when SiRE extrapolates long-term predictions from short time-series. The performance-efficiency balance and result explainability of SiRE are also validated empirically. Evaluated from the perspective of investment professionals, SiRE can precisely locate the scaleups that have a great potential return in 2 to 5 years. Furthermore, our qualitative inspection illustrates some advantageous attributes of the SiRE revenue forecasts.
\end{abstract}



\maketitle

\section{Introduction}
\label{sec:intro}

\begin{figure*}[t]
  \centering
  \includegraphics[width=.87\linewidth]{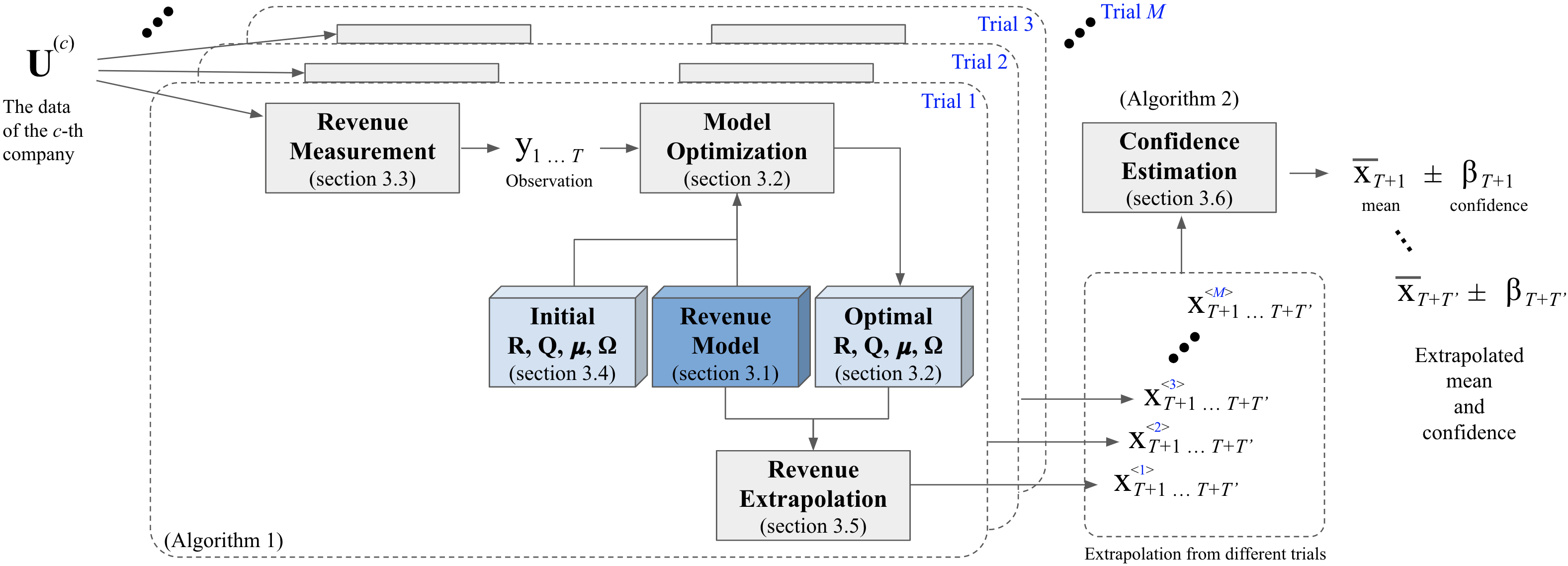}
  \vspace{-0.25cm}
  \caption{The overall procedure of SiRE: generate $T'$ revenue predictions (with confidence estimate) for the $c$-th scaleup company using $M$ trails. The corresponding sections and algorithms (pseudo in Appendix~\ref{sec:pseudo-code}) are annotated for each building block.}
\label{fig:overall}
\vspace{-0.15cm}
\end{figure*}

{\bf Revenue} is a company's total income generated from its main business activities including sales of goods and services online and offline and is commonly regarded as one of the most important performance metrics when evaluating companies. Revenue is commonly perceived as a good approximation of the company's worth, and its prediction is therefore widely considered an essential-yet-challenging component of investment analysis. 
In the public markets, the availability of large amounts of data on publicly listed companies enables investors to adopt a data-driven approach to their investment analysis. 
In contrast, data on private companies has traditionally been difficult to gather and share due to confidentiality, while data on public companies has low relevance in the private setting. 
These conditions have formed a huge obstacle for directly applying data-driven methodologies to the private capital (PC) sector, especially for smaller high-growth companies such as {\bf startups} and {\bf scaleups}. 
Startups are usually in the early stages of operations with high costs and limited revenue, meaning revenue forecast is not a must-have activity during business evaluation. 
A startup moves into scaleup territory when proving the scalability and viability of its business model and experiencing an accelerated cycle of revenue growth. 
This transition is usually accompanied by the fundraising of outside capital \cite{cavallo2019fostering}.
Consequently, revenue becomes a highly relevant metric for scaleup companies where historical values are usually available to the potential investor, even though such data remains proprietary, costly, and scarce. 
These conditions require PC investment professionals to manually and empirically forecast scaleup revenue taking some factors like business model, competitor landscape, market trends and go-to-market efficiency into consideration. 
The task is essential to evaluate the attractiveness of an investment, as it informs the change in valuation during the ownership period.
However, the level of automation, objectiveness, consistency and adaptability is far from optimal. 

To that end, it is highly desirable for PC investment professionals to have a data-driven method that performs scaleup revenue extrapolation on scarce data in an automated way. There are two usecases for such a method: (1) A quick way to assess a companies' revenue potential with little information needed; (2) benchmarking of a manually produced revenue forecasting. 
PC firms often gather proprietary data forming a small dataset $\mathbf{U}$ that contains historical revenue time-series for a number of scaleups. Without loss of generality, we assume $\mathbf{U}$ contains monthly revenue time-series:
\begin{equation}
\label{eq:dataset_U}
\begin{aligned}
\mathbf{U}=&\Big\{\!
\left(u_1^{(1)}, b_1^{(1)}, z_1^{(1)}, z_2^{(1)}\right), 
  \ldots, 
  \left(u_{T_1-1}^{(1)}, b_{T_1-1}^{(1)}, z_{T_1-1}^{(1)}, z_{T_1}^{(1)}\right), \\
  &\;\left(u_1^{(2)}, b_1^{(2)}, z_1^{(2)}, z_2^{(2)}\right), 
  \ldots, 
  \left(u_{T_2-1}^{(2)}, b_{T_2-1}^{(2)}, z_{T_2-1}^{(2)}, z_{T_2}^{(2)}\right), \\
&\;\; \cdots \; \cdots\\
&\;\left(u_1^{(C)}\!\!, b_1^{(C)}\!\!, z_1^{(C)}\!\!, z_2^{(C)}\right), 
  \ldots, 
  \left(u_{T_C-1}^{(C)}, b_{T_C-1}^{(C)}, z_{T_C-1}^{(C)}, z_{T_C}^{(C)}\right)
  \!\Big\} \\
  = & \left\{\left(u_t^{(c)}\!\!, b_t^{(c)}\!\!, z_t^{(c)}\!\!, z_{t+1}^{(c)}\right)\Big|\ c\!\in\!\mathbb{Z}\!\cap\![1,C] \wedge t\!\in\!\mathbb{Z}\!\cap\![1,T_c\!-\!1]\right\}.
\end{aligned}
\end{equation}
The tuple $\small\smash{(u_t^{(c)}, b_t^{(c)}, z_t^{(c)}, z_{t+1}^{(c)})}$ describes the revenue state for the $c$-th scaleup at the $t$-th month\footnote{For example, the tuple $\smash{(u_1^{(1)}, b_1^{(1)}, z_1^{(1)}, z_{2}^{(1)})} = (8.7, \text{Jan}\ 18, 1.5, 1.7)$ implies that company $c=1$ had a book revenue of 8.7 in January 2018, and of about 5.8 in January 2017 (as the revenue grew 1.5 times in the previous year). We can also see that the company will have increased its YoY revenue growth from 1.5 to 1.7 in February 2018, meaning the lift in revenue between February 2017 and 2018 was larger than that between January 2017 and 2018.}, where the four elements represent the book revenue (i.e. the revenues recorded in the financial records of the company), the calendar date, the current YoY (Year-over-Year) {\it revenue growth} $\small\smash{z_t^{(c)}=u_t^{(c)}/u_{t-12}^{(c)}}$, and the next YoY {\it revenue growth} $\small\smash{z_{t+1}^{(c)}=u_{t+1}^{(c)}/u_{t-11}^{(c)}}$, respectively.
For simplicity, we will omit ``YoY'' in the rest of this paper.
Variable $C$ is the total number of scaleups, and $T_c$ is the time-series length of the $c$-th scaleup.

When a new scaleup (noted as $c'$) with a historical time-series
$\small\smash{\{(u_t^{(c')}, b_t^{(c')}, z_t^{(c')}, z_{t+1}^{(c')})|t\!\in\!\mathbb{Z}\!\cap\![1,T_{c'}\!-\!1]\}}$ is evaluated,
our {\bf objective} is to extrapolate the revenue for $T'$ steps, obtaining $\small\smash{\hat{\mathfrak{x}}^{(c')}}$: 
\vspace{-0.1cm}
\begin{equation}
\label{eq:revenue_prediction_company_c_prime}
  \hat{\mathfrak{x}}^{(c')}\!=\!\left[
    \bar{x}_{T_{c'}+1}\!\pm\!\beta_{T_{c'}+1}, \;
    \bar{x}_{T_{c'}+2}\!\pm\!\beta_{T_{c'}+2}, \;
    \ldots, \;
    \bar{x}_{T_{c'}+T'}\!\pm\!\beta_{T_{c'}+T'}
  \right]
\end{equation}
where $\smash{\bar{x}_{T_{c'}+1}}$ and $\smash{\beta_{T_{c'}+t'}}$ are respectively the predicted revenue mean and error margin for time step $\smash{T_{c'}+t'}$.

In this paper, we propose an algorithm SiRE (Simulation-informed Revenue Extrapolation) that achieves the objective defined above. 
Compared to the existing methods, SiRE fulfills {\bf eight important practical requirements}: 
(1) it is sector-agnostic, which permits PC investment professionals to apply it across multiple investment cases;
(2) it works on small datasets with only a few hundred scaleups; 
(3) the extrapolation can commence from short revenue time-series, enabling revenue predictions even without granular historical data; 
(4) it should produce a fine-grained time-series of at least three-year length, a typical investment period in private markets; 
(5) each predicted revenue point should come with a confidence estimation, providing PC investment professionals with guidance on the certainty of the outcome; 
(6) it does not require any alternative data other than sector information;
(7) the model can be timely and effortlessly adapted to data change, making it seamless to integrate newly received data over time. 
(8) the prediction is explainable, promoting transparency to build trust and capture feedback.
According to our thorough literature survey (cf. Section~\ref{sec:related_work}), this is the first work that meets all practical requirements simultaneously.

\section{Related Work}
\label{sec:related_work}
Revenue time-series extrapolation for scaleups shares many commonalities with {\bf general time-series} forecasting methodologies. Existing methods for time-series forecasting can be roughly grouped into two categories: classical approaches serving as the backbone (e.g. \cite{box2015time,seeger2016bayesian,taylor2018forecasting}), and deep learning techniques following an encoder-decoder paradigm (e.g. \cite{hochreiter1997long,yu2017dilated,li2018diffusion,zhou2021informer}). Recently, the second category has become dominant, especially the Transformer-based \cite{vaswani2017attention} models such as \cite{li2019enhancing,zhou2021informer}. However, the general-purpose approaches usually require a lot of data to prevent overfitting, and they typically produce point prediction without confidence estimation. Moreover, they are not designed to exploit the common dynamics among companies in similar sectors, stages and so on. 

{\bf Financial time-series} describe the performance of companies from many different financial aspects such as sales, earnings, and stock price. Financial time-series forecasting is widely adopted by finance researchers in both academia and industry to benchmark the performance of companies.
Many machine learning (ML) based methods (e.g. \cite{rezazadeh2020generalized,uddin2021missing}) have been proposed, reporting relatively better performances compared to traditional approaches using analytical (e.g. \cite{bilinski2019analyst}) and signal processing (e.g. \cite{de2016kinetic}) techniques. In recent years, deep learning (DL) based methods have become dominant due to the greatly improved availability of data and DL frameworks. Sezer et al. \cite{sezer2020financial} carried out a thorough review (2005$\sim$2019) of DL-based financial time-series forecasting methods. 
During the past two years, the state-of-the-art DL-based methods, such as \cite{ekambaram2020attention,salinas2020deepar,wurfel2021online}, mainly adopt either RNN (Recurrent Neural Network) or Attention-based architectures \cite{vaswani2017attention}, obtaining superior results compared to ML-based baselines. 
However, the existing financial time-series forecasting methods often have one or more of the three problems: (1) dependent on alternative data beyond financial and economic sources, such as news and patent data (2) incapable of predicting long-term series, and (3) ignorant of prediction uncertainty.

{\bf Revenue} is a specific type of financial data. An accurate forecast of future revenues relies on capturing the unique dynamics of revenue development in different business sectors. Instead of using revenue time-series as input, some trials \cite{hsieh2016predicting,lee2017deep} only use aggregated financial and alternative information to predict the revenue of the very next period.
ARIMA (Autoregressive Integrated Moving Average) \cite{ariyo2014stock,box2015time} used to be a popular choice to extrapolate revenue time-series (e.g. \cite{ekmics2017revenue,valluru2018forecasting}), but it is univariate and mainly tested on single sector datasets.
Rahman et al. \cite{rahman2016revenue} attempted to produce long-term probability estimation using an exponential smoothing approach. 
In \cite{fleder2020forecasting}, a linear dynamical system (LDS) was used to impute and denoise the revenue signal. 
We have also seen some ML-based approaches such as \cite{hsieh2016predicting,zhang2020large} that usually do not provide fine-grained predictions.
The DL-based methods \cite{penpece2014predicting,mishev2019forecasting,xu2020adaptive,zhou2020domain} have started getting traction recently, illustrating strength in forecasting aggregated future revenues. 
Based on our literature survey to date, there has not been any work that simultaneously fulfill all practical requirements stated previously in Section~\ref{sec:intro}.

\section{The Proposed Approach: SiRE}

To perform revenue extrapolation with confidence estimate for scaleup companies using scarce time-series, 
we propose the SiRE approach which is visualized in Figure~\ref{fig:overall}. 
The key components and steps will be introduced in the sections that follows.

\subsection{Revenue Model}
\label{sec:revenue_model}

We assume that any revenue observed at time $t$ (i.e. $y_t$) is made up of a 
continuous and twice-differential latent component $x_t$ and a noise component $\omega_t$:
\vspace{-0.2cm}
\begin{equation}
  y_t = x_t + \omega_t,
\end{equation}
where $x_t$ is a fundamental component that captures a smooth trend of the revenue, $\omega_t$ inherits the non-differentiable external/internal noises. Inspired by \cite{de2016kinetic}, we can conveniently apply Taylor expansion (from time point $t$ to $t+\Delta t$):
\vspace{-2pt}
\begin{equation}
\label{eq:taylor_expansion}
\small
  x_{t+\Delta t} = x_t + \left.\frac{\partial x_t}{\partial t}\right|_{t=0} \cdot\Delta t
  + \frac{1}{2}\left.\frac{\partial^2 x_t}{\partial t^2}\right|_{t=0} \cdot\Delta t^2
  + \sum_{i=3}^{\infty}\frac{1}{i!}\left.\frac{\partial^i x_t}{\partial t^i}\right|_{t=0} \cdot\Delta t^i.
\end{equation}
When the value of $\Delta t$ is close to zero, we can assume the last term in Equation~\eqref{eq:taylor_expansion} is approximately zero, therefore
\begin{equation}
\label{eq:taylor_expansion_approximate}
  x_{t+\Delta t} \approx x_t + \left.\frac{\partial x_t}{\partial t}\right|_{t=0} \cdot\Delta t
  + \frac{1}{2}\left.\frac{\partial^2 x_t}{\partial t^2}\right|_{t=0} \cdot\Delta t^2,
\end{equation}
which essentially implies that the revenue at time $t+\Delta t$ can be largely calculated from the revenue at time $t$. For the sake of conciseness, we use $v_t$ and $a_t$ to denote $\smash{\left.\frac{\partial x_t}{\partial t}\right|_{t=0}}$ and $\smash{\left.\frac{\partial^2 x_t}{\partial t^2}\right|_{t=0}}$ respectively, and obtain
\begin{equation}
\label{eq:x_taylor}
  x_{t+\Delta t} \approx x_t + v_t \cdot\Delta t
  + \frac{1}{2}a_t \cdot\Delta t^2.
\end{equation}
Likewise, variable $v_t$ can be approximated with a Taylor series up to the first degree:
\vspace{-0.2cm}
\begin{equation}
\label{eq:v_taylor}
  v_{t+\Delta t} \approx v_t + a_t \cdot\Delta t, \; \text{where} \; a_t=\left.\frac{\partial^2 x_t}{\partial t^2}\right|_{t=0}=\left.\frac{\partial v_t}{\partial t}\right|_{t=0}.
\end{equation}
The terms $v_t$ and $a_t$ are analogous to the velocity and acceleration concepts in the physics world. In a stable system, we can hence assume that the acceleration term stays largely constant, i.e.:
\vspace{-3pt}
\begin{equation}
\label{eq:constant_acceleration}
 a_{t+\Delta t}\approx a_t.
\end{equation}
Of course, \eqref{eq:x_taylor} to \eqref{eq:constant_acceleration} may be expanded to derivatives of any degree, hence it is not mandatory to assume a parsimonious acceleration. 

When we try to measure a company's revenue at time $t+\Delta t$, we usually obtain a measured data point $y_{t+\Delta t}$ that constitutes the latent component $x_{t+\Delta t}$ and a systematic measurement error term from the past $d_t\cdot\Delta t$:
\vspace{-5pt}
\begin{equation}
\label{eq:measurement_dynamics}
 y_{t+\Delta t} \approx x_{t+\Delta t} + d_t\cdot\Delta t
 = (x_t + v_t \cdot\Delta t + \frac{1}{2}a_t \cdot\Delta t^2) + d_t\cdot\Delta t
\end{equation}
where $d_t$ is the unit error at time $t$ which is brought into the system by the measurement process at time $t$. We assume $d_t$ largely stays the same if $\Delta t$ is sufficiently small, thus the unit error could be scaled by $\Delta t$ and added to the latent component, yielding the measured revenue. If the same approach is utilized for measuring revenue at any time point, $d_t$ could be regarded as a relatively static term:
\vspace{-2pt}
\begin{equation}
\label{eq:static_unit_error}
 d_{t+\Delta t}\approx d_t.
\end{equation}
Since revenues are usually measured periodically (e.g. monthly and quarterly), we choose to discretize equations~\eqref{eq:x_taylor} to \eqref{eq:static_unit_error} by setting $\Delta t$=1, obtaining a linear dynamical system (LDS):
\begin{equation}
\label{eq:vectorized_dynamic_system}
  \mathbf{x}_{t+1} \approx \mathbf{A}\mathbf{x}_t
  \;\; \text{and} \;\;
  y_{t} \approx \mathbf{c} \mathbf{x}_t,
  \;\; \text{{where}} \;
  \mathbf{x}_t=[y_t, x_t, v_t, a_t, d_t]^\top.
\end{equation}
Vector $\mathbf{x}_t$ is the latent state vector for the $t$-th time step.
Vector $\mathbf{c}$ is responsible for transforming the latent state to the measurable revenue, hence $\mathbf{c}$ is called measurement vector. 
Matrix $\mathbf{A}$ transforms the current latent state to the next one.
According to equations~\eqref{eq:x_taylor} to \eqref{eq:static_unit_error}, matrices $\mathbf{A}$ and $\mathbf{c}$ should take the values of 
\begin{equation}
\label{eq:matrix_A_c}
  \renewcommand*{\arraystretch}{0.5}
  \mathbf{A}=
  \begin{bmatrix}
    0 & 1 & 1 & 1/2 & 1 \\
    0 & 1 & 1 & 1/2 & 0 \\
    0 & 0 & 1 & 1 & 0 \\
    0 & 0 & 0 & 1 & 0 \\
    0 & 0 & 0 & 0 & 1
  \end{bmatrix}
  \;\; \text{and} \;\;
  \mathbf{c}=[1,0,0,0,0],
\end{equation}
To make the right-hand-side terms exactly equivalent to the left-hand-side ones in equations~\eqref{eq:vectorized_dynamic_system}, we add time-dependent noise terms $\boldsymbol{\omega}_t$ and $\boldsymbol{\nu}_t$ to both equations, getting a more general form of
\begin{equation}
\label{eq:lds_dynamics}
  \mathbf{x}_{t+1} = \mathbf{A}\mathbf{x}_t + \boldsymbol{\omega}_t 
  \;\; \text{and} \;\;\;
  y_{t} = \mathbf{c} \mathbf{x}_t + \boldsymbol{\epsilon}_t,
\end{equation}
where $\boldsymbol{\omega}_t$ and $\boldsymbol{\epsilon}_t$ are zero-mean normally-distributed random variables with covariance matrices  $\mathbf{Q}$ and $\mathbf{R}$, i.e. $\boldsymbol{\omega}_t\sim\mathcal{N}(0, \mathbf{Q})$ and $\boldsymbol{\epsilon}_t\sim\mathcal{N}(0, \mathbf{R})$. As a result, the company revenue can be modeled as a linear time-variant dynamical system, also known as linear Gaussian state-space models. The measured revenue $\mathbf{y}_{t}$ is a linear function of the state, $\mathbf{x}_t$, and the state at any time step depends linearly on the previous state.  


\subsection{Model Optimization}
\label{sec:model_training}
The revenue model unfolds recursively from the initial state $\mathbf{x}_1$, which is a normal random vector with mean vector $\boldsymbol{\mu}\in\mathbb{R}^5$ and a $5\times 5$ covariance matrix $\boldsymbol{\Omega}$. 
The optimizable parameters of the model are $\mathbf{Q}$, $\mathbf{R}$, $\boldsymbol{\mu}$, and $\boldsymbol{\Omega}$. 
The goal is to maximize $\small\smash{\mathfrak{O} = \mathbb{E}[\log p(\{\mathbf{x}\},\{y\})\vert\{y\}]}$, which is the joint log likelihood of $\{y\}=(y_1,y_2,\ldots,y_T)$ and $\{\mathbf{x}\}=(x_1,x_1,\ldots,x_T)$, conditioned on the observable $\{y\}$:
\begin{equation}
\label{eq:conditional_log_likelihood}
\small
\begin{aligned}
&\begin{aligned}
\mathfrak{O} = & \;\mathbb{E}[\log p(\{\mathbf{x}\},\{y\})\vert\{y\}] \\
  = & -\frac{1}{2}\Tr\left\{\mathbf{R}^{-1}\textstyle{\sum}_{t=1}^T \left[(y_t-\mathbf{c}\mathbf{x}_t^T)^2
  +\mathbf{c}\mathbf{P}_t^T\mathbf{c}^\top
   \right]\right\}
  - \frac{T}{2}\log\vert \mathbf{R}\vert \\
  & -\frac{1}{2}\Tr\left\{\mathbf{Q}^{-1}(\mathbf{G}-\mathbf{F}\mathbf{A}^{\top}\!-\!\mathbf{A}\mathbf{F}^{\top}+\mathbf{A}\mathbf{E}\mathbf{A}^{\top})\right\}
  \!-\! \frac{T\!-\!1}{2}\log\vert \mathbf{Q}\vert \\
  & - \frac{1}{2}\Tr\left\{\boldsymbol{\Omega}^{-1}\left[\mathbf{P}_1^T+(\mathbf{x}_1^T-\boldsymbol{\mu})(\mathbf{x}_1^T-\boldsymbol{\mu})^\top\right]\right\}
  - \frac{1}{2}\log\vert \boldsymbol{\Omega}\vert,
\end{aligned} \\
&\begin{aligned}
  \text{where }\mathbf{E}= & \textstyle{\sum}_{t=2}^T\left[\mathbf{P}_{t-1}^T+\mathbf{x}_{t-1}^T(\mathbf{x}_{t-1}^T)^\top\right], \\
  \mathbf{G}= & \textstyle{\sum}_{t=1}^T\left[\mathbf{P}_t^T+\mathbf{x}_t^T(\mathbf{x}_t^T)^\top\right]\; \text{and} \\
  \mathbf{F}= & \textstyle{\sum}_{t=2}^T\left[\mathbf{P}_{t,t-1}^T+\mathbf{x}_t^T(\mathbf{x}_{t-1}^T)^\top\right].
\end{aligned}
\end{aligned}
\end{equation}
The operation $\Tr(\cdot)$ denotes the trace calculation, $\smash{\{y\}_{t_0}^{t_1}}$ represents a sub-sequence of $\{y\}$, i.e. $\small\smash{\{y\}_{t_0}^{t_1}=(y_{t_0}, y_{t_0+1}, \ldots, y_{t_1})}$, and $\smash{\mathbf{x}_t^\tau}$ defines the conditional mean $\small\smash{\mathbb{E}(\mathbf{x}_t\vert \{y\}_1^\tau)}$. 
The terms $\smash{\mathbf{P}_t^{\tau}}$ and $\smash{\mathbf{P}_{t, t-1}^{\tau}}$ define the covariances $\smash{\Cov(\mathbf{x}_t\vert\{y\}_1^\tau)}$ and $\smash{\Cov(\mathbf{x}_t\mathbf{x}_{t-1}\vert\{y\}_1^\tau)}$, respectively.
The derivation of \eqref{eq:conditional_log_likelihood} can be found in Appendix~\ref{sec:derive_opt_target}.
To maximize \eqref{eq:conditional_log_likelihood}, we adopt the EM (Expectation Maximization)\footnote{EM algorithm is used due to its universality and popularity. There are analytical solutions (e.g. \cite{huang2020analyzing}) that may give more stable parameter estimations.} algorithm \cite{dempster1977maximum,byron2004derivation} explained in Appendix~\ref{sec:em-detail}.
The convergence properties of EM are discussed in \cite{wu1983convergence}.


\begin{figure}[t]
  \centering
  \includegraphics[width=\linewidth]{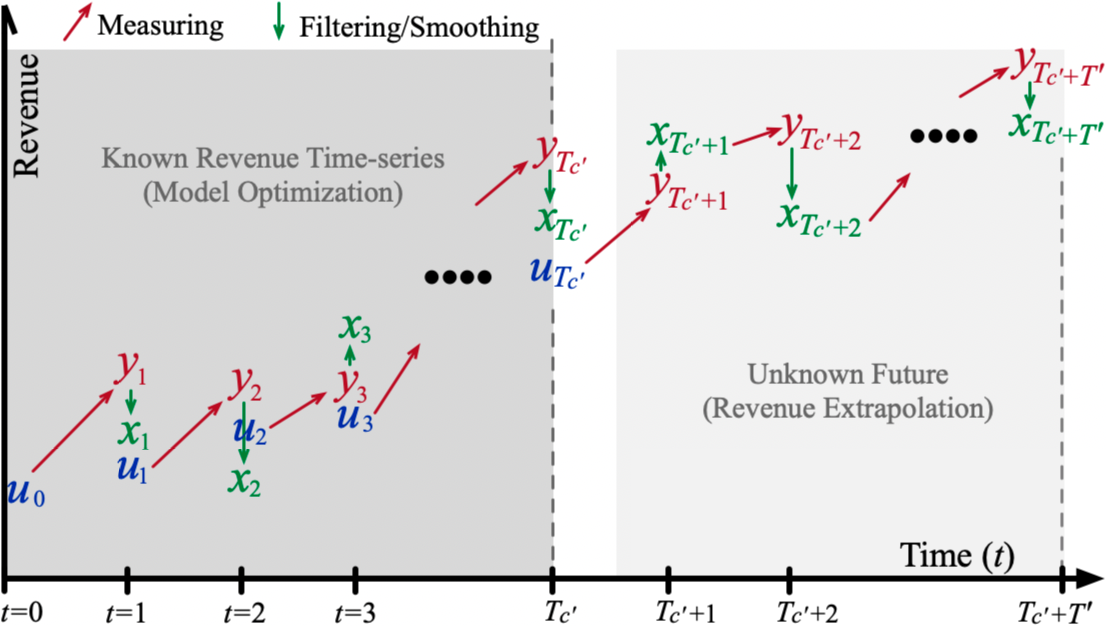}
  \vspace{-0.65cm}
  \caption{A high-level illustration of the optimization and extrapolation process that comprises measuring and filtering/smoothing operations. During optimization, the booked revenue $u_t$ is used as the measuring base (the starting positions of red arrows); but in extrapolation, the filtered revenue $x_t$ is used due to the unavailability of $u_t$.}
  \Description{Need to create a proper one later, this one very much just a sketch.}
\label{fig:measuring_revenue}
\vspace{-0.2cm}
\end{figure}

\subsection{Revenue Measurements}
\label{sec:revenue_measurement}
Obtaining measurements of the revenue signal is crucial (cf. formulations in \ref{sec:model_training}).
One might argue to simply use the data from financial bookings as the measurement of revenue, but this type of revenue measurement is not available in the future.
Appendix~\ref{sec:whynot} conducts an extensive discussion on why booked revenue should not be directly used as measurement. 
We propose a simulation-informed approach, illustrated in Figure~\ref{fig:measuring_revenue}, that enables measuring the revenue in both past and future time horizons.
Specifically, to measure the revenue for a company-in-focus $c'\!\in\![1,C]$ at time point $t'\!+\!1$, 
we assume the approximated revenue signal from the previous time point $t'\!>\!0$ is known: the book revenue $\smash{u_{t'}^{(c')}}$ (for the past known period: left part of Figure~\ref{fig:measuring_revenue}) or latent state $\small\smash{x_{t'}^{(c')}}$ (for the future unknown period: right part of Figure~\ref{fig:measuring_revenue}).
In this section, we mainly focus on measuring revenues in the past, while the measuring process for the future will be detailed in Section~\ref{sec:model_forecasting}.
The objective here is to measure the noisy value of the current revenue $\small\smash{y_{t'+1}^{(c')}}$. The overall approach is to 
(1) construct a {\it measuring dataset} $\small\overline{\mathbf{U}}\subset\mathbf{U}$ that contains similar states to company $c'$ at time point $t'$, 
and (2) calculate $\small\smash{y_{t'+1}^{(c')}}$ by sampling a likely revenue growth from $\overline{\mathbf{U}}$.

\subsubsection{Construct measuring dataset}
\label{sec:measuring_dataset_train}
The measuring dataset is obtained by applying a sequence of filtering operations.

{\bf Filter by business.} Each company has its main operating sector (e.g. \verb|Fintech| and \verb|Gaming|) and customer focus (e.g. \verb|B2B| and \verb|B2C|). We use $S_{c}$ to denote the concatenation of sector and customer focus for company $c$. For example, $S_{c'}$=``\verb|Fintech+B2C|'' indicates that company $c'$ belongs to the \verb|Fintech| sector, and has a \verb|B2C| customer focus. For the sake of simplicity, we assume $S_{c'}$ does not change during the life span of company $c'$. In $\mathbf{U}$, {\it we only keep the companies with the same sector and customer focus as the company-in-focus}, i.e. $S_c=S_{c'}$ as shown in \eqref{eq:dataset_measuring_subset}.

{\bf Filter by date.} It is common sense that when one tries to estimate the current revenue, {\it there is no way to peek into the future for ground-truth information}. Likewise, to obtain $y_{t'+1}$ (for date $\small\smash{b_{t'+1}^{(c')}}$) knowing $\small\smash{u_{t'}^{(c')}}$ (for date $\small\smash{b_{t'}^{(c')}}$), we should avoid using any data that comes after $\small\smash{b_{t'}^{(c')}}$. 
Therefore, we need to apply the filter $\small\smash{b_t^{(c)}\leq b_{t'}^{(c')}}$ on dataset $\mathbf{U}$, where the calendar date ${\small b_{t'}^{(c')}}$ is called the cutoff date, splitting the timeline into pseudo ``past'' and ``future'' periods.

{\bf Filter by revenue.} A na\"ive way of estimating revenue is to look at how other scaleup companies have developed from a similar stage, and {\it the stage of a scaleup could be approximated by the scale of its latest known revenue}. 
Generally, to obtain $\small\smash{y_{t'+1}^{(c')}}$, the latest known revenue is $\small\smash{u_{t'}^{(c')}}$. In $\mathbf{U}$, we only keep the tuples satisfying $\small\smash{u_t^{(c)}\in[(1-r)u_{t'}^{(c')}, (1+r)u_{t'}^{(c')}]}$, where $r\in(0,1)$ is a relaxing term that controls the size of the measuring dataset. 
Putting all filtering operations above together, we obtain a set $\small\smash{\mathbf{U}_{t'}^{(c')}}$ defined as
\begin{equation}
\label{eq:dataset_measuring_subset}
\small
\mathbf{U}_{t'}^{(c')}\!\!\coloneqq\!
\left\{
    \left(u_t^{(c)}, b_t^{(c)}, z_t^{(c)}, z_{t+1}^{(c)}\right)
\;\middle\vert\;
\begin{aligned}
  & \left(u_t^{(c)}, b_t^{(c)}, z_t^{(c)}, z_{t+1}^{(c)}\right) \in \mathbf{U} \\
  & \wedge\; S_c=S_{c'} \wedge\; b_t^{(c)}\leq b_{t'}^{(c')}  \;\wedge\\
  & u_t^{(c)}\!\!\!\in\!\!\left[(1\!-\!r)u_{t'}^{(c')}, (1\!+\!r)u_{t'}^{(c')}\right]
\end{aligned}
\right\}.
\end{equation}

{\bf Filter by revenue growth.} Besides the current revenue, {\it the latest known revenue growth $\small\smash{z_{t'}^{(c')}}$can also indicate the development stage of a scaleup}. To start with, we use $\grave{\mathbf{Z}}$ to denote the set containing all $\small\smash{z_t^{(c)}}$ values that appear in
$\small\smash{\mathbf{U}_{t'}^{(c')}}$:
\begin{equation}
\label{eq:Z_in_measuring_subset}
\grave{\mathbf{Z}} \coloneqq 
\left\{
    z_t^{(c)}
\;\middle\vert\;
    \forall \left(u_t^{(c)}, b_t^{(c)}, z_t^{(c)}, z_{t+1}^{(c)}\right) \in \mathbf{U}_{t'}^{(c')}
\right\}.
\end{equation}
We divide $\small\smash{\grave{\mathbf{Z}}}$ evenly into $n$ quantiles, 
obtaining $n+1$ quantile boundaries $\small\smash{q_0, q_1, \ldots, q_{n-1}, q_n}$, where $\smash{q_0=\min(\grave{\mathbf{Z}})}$ and $\smash{q_n=\max(\grave{\mathbf{Z}})}$. Assume $\small\smash{z_{t'}^{(c')}}$ falls in the $k$-th quantile, i.e. $\small\smash{z_{t'}^{(c')}\in[q_{k-1}, q_k)}$, we gather the tuples (from set $\small\smash{\mathbf{U}_{t'}^{(c')}}$) with $\smash{z_t^{(c)}}$ in the same range, forming the {\it measuring dataset} $\overline{\mathbf{U}}$:
\begin{equation}
\label{eq:dataset_growth_subset}
\small
\begin{aligned}
& \overline{\mathbf{U}} \coloneqq 
\left\{
    \left(u_t^{(c)}, b_t^{(c)}, z_t^{(c)}, z_{t+1}^{(c)}\right) 
\;\middle\vert\;
\begin{aligned}
  & \left(u_t^{(c)}, b_t^{(c)}, z_t^{(c)}, z_{t+1}^{(c)}\right) \in \mathbf{U}_{t'}^{(c')} \\ & \;\wedge\; z_t^{(c)}\in[q_{k-1}, q_k)
\end{aligned}
\right\} \\
& \text{\bf{s.t.}}\;z_{t'}^{(c')}\in[q_{k-1}, q_k), \; k\in\mathbb{Z}\cap[1,n],
\end{aligned}
\end{equation}
where the tuples $\smash{(u_t^{(c)}, b_t^{(c)}, z_t^{(c)}, z_{t+1}^{(c)})}$ can be viewed as known benchmarking snapshots of revenue dynamics that are comparable to the scaleup-in-focus $c'$ at date $\smash{b_{t'}^{(c')}}$. The total number of quantiles $n$ needs to be searched, and we empirically find that $n=4$ gives reasonable revenue measurements.

\subsubsection{Measure revenue by sampling a growth}
\label{sec:measure_rev_sample_growth}
Recall that we want to measure the revenue $\small\smash{y_{t'+1}^{(c')}}$, which is equivalent to calculating the revenue growth $\small\smash{z_{t'+1}^{(c')}}$ since $\smash{y_{t'+1}^{(c')}\approx u_{t'}^{(c')}\cdot (z_{t'+1}^{(c')})^{1/12}}$ where $\small\smash{u_{t'}^{(c')}}$ is the known previous revenue.
Because $\smash{\overline{\mathbf{U}}}$ contains similar states to company $c'$ at time point $t'$, we may sample the last element (i.e. $\smash{z_{t+1}^{(c)}}$) of tuples from $\smash{\mathbf{U}_{t'}^{(c')}}$. 
As a result, we extract all $\smash{z_{t+1}^{(c)}}$ from $\smash{\overline{\mathbf{U}}}$:
\begin{equation}
\label{eq:Z_next_in_U_bar}
\acute{\mathbf{Z}} \coloneqq 
\left\{
    z_{t+1}^{(c)}
\;\middle\vert\;
    \forall \left(u_t^{(c)}, b_t^{(c)}, z_t^{(c)}, z_{t+1}^{(c)}\right) \in \overline{\mathbf{U}}
\right\},
\end{equation}
where $\acute{\mathbf{Z}}$ contains the future revenue growth rates that once occurred to scaleups in a state comparable to scaleup $c'$ at time $t'$. Now, we sample one value from the underlying distribution of $\acute{\mathbf{Z}}$ using a two-step procedure:
\vspace{-0.3cm}
\begin{equation}
\label{eq:sample_next_growth}
\text{first} \;\;
\acute{z} \sim \mathcal{U}(\acute{\mathbf{Z}})
\;,\;\; \text{then} \;\;
\hat{z} \sim \mathcal{N}\left(\acute{z}, \;
  \left[\frac{4}{3}\cdot\frac{\sigma(\acute{\mathbf{Z}})^5}{|\acute{\mathbf{Z}}|}\right]^{\frac{1}{5}}\right),
\end{equation}
where $|\acute{\mathbf{Z}}|$ denotes the number of elements in set $\acute{\mathbf{Z}}$ and $\sigma(\acute{\mathbf{Z}})$ represents the standard deviation over $\acute{\mathbf{Z}}$.
Equation~\eqref{eq:sample_next_growth} first samples a value $\acute{z}$ uniformly from set $\acute{\mathbf{Z}}$, then it samples a value $\hat{z}$ from a $\acute{z}$-mean Normal distribution with a variance calculated following the Silverman's rule of thumb \cite{silverman2018density}. Per revenue growth definition, we approximate $\small\smash{y_{t'+1}^{(c')}}$ with
\begin{equation}
\label{eq:final_measurement}
y_{t'+1}^{(c')} \coloneqq u_{t'}^{(c')} \cdot \hat{z}^{\frac{1}{12}}.
\end{equation}

\subsection{Parameter Initialization}
\label{sec:model_initialization}
As discussed previously, the parameters that need to be optimized are $\mathbf{Q}$, $\mathbf{R}$, $\boldsymbol{\mu}$, and $\boldsymbol{\Omega}$, hence we need to assign reasonable initial values to them. For $\mathbf{Q}$, $\mathbf{R}$, and $\boldsymbol{\Omega}$, we simply set
\begin{equation}
\label{eq:init_Q_R_Omega}
\mathbf{Q}=\boldsymbol{\Omega}=\mathbf{I}_5 \;\;\; \text{and} \;\;\;  \mathbf{R}=\mathbf{I}_1.
\end{equation}

For any company, we know its booked revenue $u_0, u_1, \ldots, u_T$. Following the measuring procedure described in Section~\ref{sec:revenue_measurement}, we can obtain $y_1, y_2, \ldots, y_T, y_{T+1}$, and for any $t\in\mathbb{Z}\cap[1,T]$ we define $d_t=y_t-u_t$, obtaining $d_1, d_2, \ldots, d_T$. 
Recalling that $\boldsymbol{\mu}$ is the mean vector of the initial state $\mathbf{x}_1=[y_1,x_1,v_1,a_1,d_1]^\top$, we can intuitively define $\boldsymbol{\mu}$ as
\begin{equation}
\label{eq:init_mu}
  \boldsymbol{\mu} \coloneqq
  \begin{bmatrix}
    u_0 + \frac{1}{T}\sum_{t=1}^T d_t  \\
    u_0  \\
    (u_1 - u_0) / \Delta t  \\
    [(u_2 - u_1) - (u_1 - u_0)] / (2\Delta t)  \\
    \frac{1}{T}\sum_{t=1}^T d_t 
  \end{bmatrix}, \; \text{where} \; \Delta t=1.
\end{equation}

\subsection{Revenue Extrapolation}
\label{sec:model_forecasting}
After we have trained our revenue model (Section~\ref{sec:revenue_model}) with historical measurements (Section~\ref{sec:revenue_measurement}) and initial parameters (Section~\ref{sec:model_initialization}) for the $c'$-th company, we can start extrapolating the revenue into the future using the measurements $\small\smash{y_{T_{c'}+1}^{(c')}}$, where $T_{c'}$ is the total number of available book revenues for company $c'$. 
Forecasting $T'$ revenues into the future is equivalent to estimating $\small\smash{x_{\tau}^{(c')}}$, where $\small\smash{\tau\in\mathbb{Z}\cap[T_{c'}+1,T_{c'}+T']}$. 
Notice that we already calculated $\smash{y_{T_{c'}+1}^{(c')}}$ in Section~\ref{sec:revenue_measurement}, so the first predicted revenue $\small\smash{x_{T_{c'}+1}^{(c')}}$ can be extracted from the vector of $\small\smash{\mathbf{x}_{T_{c'}+1}^{(c')}}$ (i.e. $\small\smash{\mathbf{x}_{T_{c'}+1}^{T_{c'}+1}}$) obtained via forward filtering following Equation~\eqref{eq:forward_filtering} in Appendix~\ref{sec:em-detail}. 
Afterwards, we repeat a two-step operation (i.e. \ref{sec:measuring_in_forecasting} measuring and \ref{sec:filtering_in_forecasting} filtering) for $T'\!-\!1$ times so that we obtain $T'$ predicted revenues. 
Figure~\ref{fig:measuring_revenue} gives an illustration of this process.
In the end, we apply a global smoothing (Section~\ref{sec:global_smoothing_in_forecasting}) to nudge the revenues using all measurements (historical and future).

\subsubsection{The measuring step}
\label{sec:measuring_in_forecasting}
When extrapolating for scaleup company $c'$, we know the filtered revenues $\smash{x_{\tau}^{(c')}}$ but not the booked ones $\smash{u_{\tau}^{(c')}}$. Therefore, as shown in the right part of Figure~\ref{fig:measuring_revenue}, we use $\smash{x_{\tau}^{(c')}}$ as an approximation of $\smash{u_{\tau}^{(c')}}$. 
At the $\tau+1$ prediction step, we know the values of $\smash{x_{\tau}^{(c')}}$, and the corresponding revenue growth $\smash{z_{\tau}^{(c')}}$ can be simply estimated following their definition:
\begin{equation}
\label{eq:revenue_growth_in_prediction}
\begin{aligned}
& z_{\tau}^{(c')} = 
\begin{cases}
x_{\tau}^{(c')} / x_{\tau -12}^{(c')} & , \text{if}\;\; u_{\tau -12}^{(c')} \;\;\text{is not available}\\
x_{\tau}^{(c')} / u_{\tau -12}^{(c')} & , \text{if}\;\; u_{\tau -12}^{(c')} \;\;\text{is available}
\end{cases}, \\ 
& \text{where } \tau\in\mathbb{Z}\cap[T_{c'}+1,T_{c'}+T'].
\end{aligned}
\end{equation}
The measuring operation aims at obtaining $\smash{y_{\tau}^{(c')}}$ using an approach described in Section~\ref{sec:revenue_measurement} with a few minor differences. Like Equation~\eqref{eq:dataset_measuring_subset}, we first obtain set $\smash{\mathbf{U}_{\tau}^{(c')}}$.
\begin{equation}
\label{eq:forecast_measuring_dataset}
\small
\mathbf{U}_{\tau}^{(c')} \coloneqq
\left\{
    \left(u_t^{(c)}, b_t^{(c)}, z_t^{(c)}, z_{t+1}^{(c)}\right)
\;\middle\vert\;
\begin{aligned}
  & \left(u_t^{(c)}, b_t^{(c)}, z_t^{(c)}, z_{t+1}^{(c)}\right) \in \mathbf{U} \\
  & \;\wedge\; S_c=S_{c'} \wedge\; b_t^{(c)}\leq b_{\tau}^{(c')} \;\wedge\\
  & u_t^{(c)}\!\!\!\in\!\!\left[\!(1\!-\!r)u_{\tau}^{(c')}, (1\!+\!r)u_{\tau}^{(c')}\!\right]
\end{aligned}
\right\},
\end{equation}
where $\smash{b_{\tau}^{(c')}}$ is the mapped calendar date of $\tau$ for company $c'$. 
Similar to Equation~\eqref{eq:Z_in_measuring_subset}, we extract set $\grave{\mathbf{Z}}$ from $\smash{\mathbf{U}_{\tau}^{(c')}}$. 
We then find the boundaries of $n$ quantiles over set $\smash{\grave{\mathbf{Z}}}$: $\smash{q_0, q_1, \ldots, q_{n-1}, q_n}$, 
and form the measuring dataset $\overline{\mathbf{U}}'$ for prediction:
\begin{equation}
\label{eq:forecast_growth_subset}
\small
\begin{aligned}
& \overline{\mathbf{U}}' \coloneqq 
\left\{
    \left(u_t^{(c)}, b_t^{(c)}, z_t^{(c)}, z_{t+1}^{(c)}\right) 
\;\middle\vert\;
\begin{aligned}
  & \left(u_t^{(c)}, b_t^{(c)}, z_t^{(c)}, z_{t+1}^{(c)}\right) \in \mathbf{U}_{\tau}^{(c')} \\ & \;\wedge\; z_t^{(c)}\in[q_{k-1}, q_k)
\end{aligned}
\right\} \\
& \text{\bf{s.t.}}\;z_{\tau}^{(c')}\in[q_{k-1}, q_k), \; k\in\mathbb{Z}\cap[1,n],
\end{aligned}
\end{equation}
where parameter $n$ should use exactly the same value as in Equation~\eqref{eq:dataset_growth_subset}. Thereafter, we sample a revenue growth $\hat{z}$ by strictly following Equation~\eqref{eq:Z_next_in_U_bar} and \eqref{eq:sample_next_growth}. Finally, we compute $\small\smash{y_{\tau+1}^{(c')}}$ with
\begin{equation}
\label{eq:forecast_final_measurement}
y_{\tau+1}^{(c')} \coloneqq x_{\tau}^{(c')} \cdot \hat{z}^{\frac{1}{12}},
\; \text{where} \;  
\tau\in\mathbb{Z}\cap[T_{c'}+1,T_{c'}+T'-1].
\end{equation}

\subsubsection{The filtering step}
\label{sec:filtering_in_forecasting}
Now that we know the values of $\small\smash{y_{\tau+1}^{(c')}}$ and $\small\smash{\mathbf{x}_{\tau}^{(c')}}$, we can perform the forward filtering procedure, as introduced in Appendix~\ref{sec:em-forward-filtering}, to obtain $\small\smash{\mathbf{x}_{\tau+1}^{(c')}}$:
\begin{equation}
\label{eq:forecast_filtering}
\begin{aligned}
  \mathbf{x}_{\tau+1}^{\tau} = & \mathbf{A}\mathbf{x}_{\tau}^{\tau}\;, \\
  \mathbf{P}_{\tau+1}^{\tau} = & \mathbf{A}\mathbf{x}_{\tau}^{\tau}\mathbf{A}^\top+\mathbf{Q}\;, \\
  \mathbf{K}_{\tau+1} = & \mathbf{P}_{\tau+1}^{\tau}\mathbf{c}^\top(\mathbf{c}\mathbf{P}_{\tau+1}^{\tau}\mathbf{c}^\top + \mathbf{R})^{-1}\;\;, \\
  \mathbf{P}_{\tau+1}^{\tau+1} = & \mathbf{P}_{\tau+1}^{\tau} - \mathbf{K}_{\tau+1}\mathbf{c}\mathbf{P}_{\tau+1}^{\tau}\;, \\
  \mathbf{x}_{\tau+1}^{\tau+1} = & \mathbf{x}_{\tau+1}^{\tau} + \mathbf{K}_{\tau+1}(y_{\tau+1} - \mathbf{c}\mathbf{x}_{\tau+1}^{\tau})\;,
\end{aligned}
\end{equation}
which is essentially the same as Equation~\eqref{eq:forward_filtering} in Appendix~\ref{sec:em-detail} with minor changes.
Notations $\small\smash{y_{\tau+1}}$, $\small\smash{\mathbf{x}_{\tau}^{\tau}}$ and $\small\smash{\mathbf{x}_{\tau+1}^{\tau+1}}$ are respectively equivalent to $\small\smash{y_{\tau+1}^{(c')}}$, $\small\smash{\mathbf{x}_{\tau}^{(c')}}$ and $\small\smash{\mathbf{x}_{\tau+1}^{(c')}}$ when the company identifier $c'$ is neglected. 
By definition, we can take the second element (denoted $\small\smash{x_{\tau+1}^{c'}}$) from $\small\smash{\mathbf{x}_{\tau+1}^{\tau+1}}$ in order to calculate $\small\smash{z_{\tau+1}^{(c')}}$ when starting to measure the next revenue in Equation~\eqref{eq:revenue_growth_in_prediction}.

\subsubsection{The global smoothing}
\label{sec:global_smoothing_in_forecasting}
We iterate the above measuring and filtering steps $T'-1$ times, 
obtaining $\small\smash{\mathbf{x}_{\tau+1}^{\tau+1}}$, $\small\smash{\mathbf{P}_{\tau+1}^{\tau}}$ and $\small\smash{\mathbf{P}_{\tau+1}^{\tau+1}}$, where $\smash{\tau\in\mathbb{Z}\cap[T_{c'}+1,T_{c'}+T'-1]}$. 
In the beginning of forecasting, 
we obtained $\small\mathbf{x}_{T_{c'}+1}^{T_{c'}+1}$, 
$\small\smash{\mathbf{P}_{T_{c'}+1}^{T_{c'}}}$ 
and $\small\smash{\mathbf{P}_{T_{c'}+1}^{T_{c'}+1}}$. 
When the model was optimized in Section~\ref{sec:model_training} (cf. Appendix~\ref{sec:em-detail}), we also obtained $\small\smash{\mathbf{x}_{t}^{t}}$, $\mathbf{P}_{t}^{t-1}$, and $\small\mathbf{P}_{t}^{t}$, where  $\smash{t\in\mathbb{Z}\cap[1,T_{c'}]}$. 
To unify, if we define $\smash{\tilde{t}\in\mathbb{Z}\cap[1,T_{c'}+T']}$, then we know all the values of $\small\smash{\mathbf{x}_{\tilde{t}}^{\tilde{t}}}$, $\small\smash{\mathbf{P}_{\tilde{t}}^{\tilde{t}-1}}$, and $\small\smash{\mathbf{P}_{\tilde{t}}^{\tilde{t}}}$. It is viable to smooth our predictions using all measurements so that the resulting revenue predictions are correlated better among one another. Concretely, we apply the following recursively from $\tilde{t}\!=\!T_{c'}+T'$ all the way till $\tilde{t}=1$.
\begin{equation}
\label{eq:global_smoothing_in_forecasting}
\begin{aligned}
  \mathbf{J}_{\tilde{t}-1} = & \mathbf{P}_{\tilde{t}-1}^{\tilde{t}-1}\mathbf{A}^\top(\mathbf{P}_{\tilde{t}}^{\tilde{t}-1})^{-1}\;, \\
  \mathbf{x}_{\tilde{t}-1}^{T_{c'}+T'} = & \mathbf{x}_{\tilde{t}-1}^{\tilde{t}-1} + \mathbf{J}_{\tilde{t}-1}(\mathbf{x}_{\tilde{t}}^{T_{c'}+T'} - \mathbf{A}\mathbf{x}_{\tilde{t}-1}^{\tilde{t}-1})\;, \\
  \mathbf{P}_{\tilde{t}-1}^{T_{c'}+T'} = & \mathbf{P}_{\tilde{t}-1}^{\tilde{t}-1} + \mathbf{J}_{\tilde{t}-1}(\mathbf{P}_{\tilde{t}}^{T_{c'}+T'} - \mathbf{P}_{\tilde{t}-1}^{\tilde{t}-1})\mathbf{J}_{\tilde{t}-1}^\top\;.
\end{aligned}
\end{equation}
To this point, we manage to obtain $\small\mathbf{x}_{T_{c'}+1}^{T_{c'}+T'}, \mathbf{x}_{T_{c'}+2}^{T_{c'}+T'}, \ldots, \mathbf{x}_{T_{c'}+T'}^{T_{c'}+T'}$.
According to the definition in Equation \eqref{eq:vectorized_dynamic_system}, the final revenue forecasts $\small\smash{x_{T_{c'}+1}^{(c')}, x_{T_{c'}+2}^{(c')}, \ldots, x_{T_{c'}+T'}^{(c')}}$ (i.e. $x_{T_{c'}+1}, x_{T_{c'}+2}, \ldots, x_{T_{c'}+T'}$ ) are simply the second elements in those corresponding state vectors:
\begin{equation}
\label{eq:final_revenue_forecasts}
\left[
x_{T_{c'}+1}, \ldots, x_{T_{c'}+T'}
\right]
=
\left[\left(\mathbf{x}_{T_{c'}+1}^{T_{c'}+T'}\right)_1, \ldots,
\left(\mathbf{x}_{T_{c'}+T'}^{T_{c'}+T'}\right)_1\right],
\end{equation}
where operation $(\cdot)_1$ returns the second element (with index starting from zero) from a vector.

\begin{figure}[t]
  \centering
  \includegraphics[width=\linewidth]{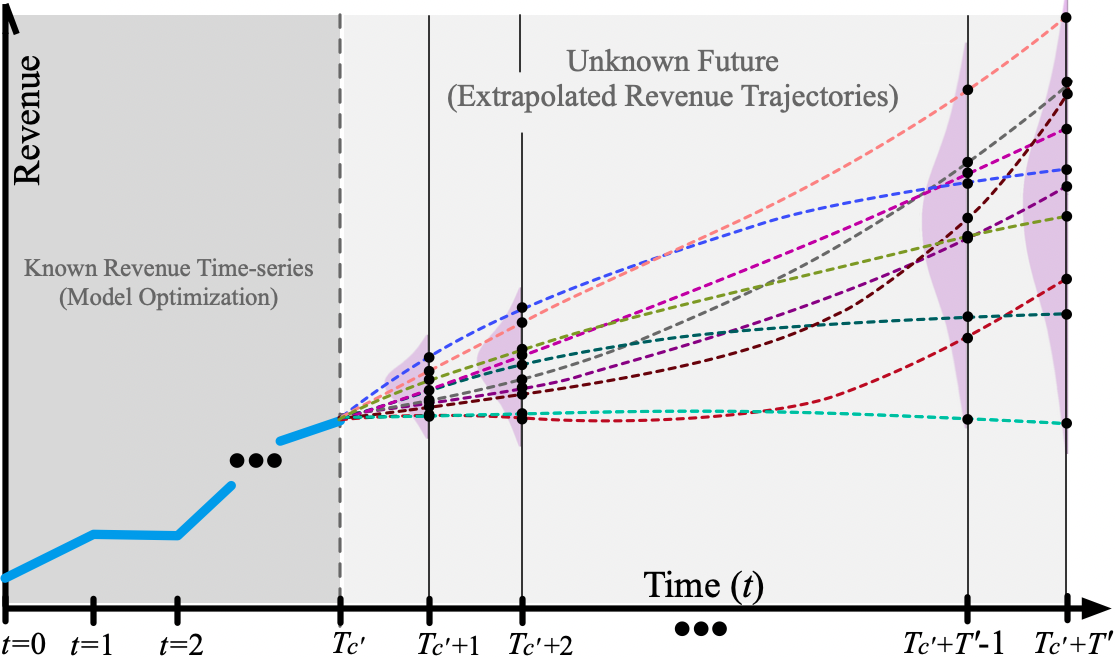}
  \vspace{-0.6cm}
  \caption{Illustration of the underlying normal distribution of the predicted revenue for any future time point $T_{c'}+t'$.}
\label{fig:revenue_distribution}
\end{figure}

\begin{table*}[t]
\small
\addtolength{\tabcolsep}{3.2pt}
\renewcommand{\arraystretch}{0.9}
\centering
\begin{tabular}{c|c|rr|rr|rr|rr|rr}
\bottomrule
\multicolumn{2}{c|}{\textbf{Metrics}:} & \multicolumn{2}{c|}{\textbf{RMSE}} & \multicolumn{2}{c|}{\textbf{MAPE}} & \multicolumn{2}{c|}{\textbf{PCC}} & \multicolumn{2}{c|}{\textbf{NLL}} & \multicolumn{2}{c}{\textbf{ACC}} \\
\hline
\multicolumn{2}{c|}{Dataset:} & \multicolumn{1}{c}{ARR129} & \multicolumn{1}{c|}{SapiQ} & \multicolumn{1}{c}{ARR129} & \multicolumn{1}{c|}{SapiQ} & \multicolumn{1}{c}{ARR129} & \multicolumn{1}{c|}{SapiQ} & \multicolumn{1}{c}{ARR129} & \multicolumn{1}{c|}{SapiQ} & \multicolumn{1}{c}{ARR129} & \multicolumn{1}{c}{SapiQ} \\
\hline
\parbox[t]{2mm}{\multirow{6}{*}{\rotatebox[origin=c]{90}{Methods}}} 
& SiRE (Ours) & {\bf 9.6917} & {\bf 57.8620} & {\bf 0.0480} & {\bf 0.6571} & {\bf 0.8284} & {\bf 0.6049} & {\bf 7.0578} & {\bf 8.5866} & {\bf 0.7102} & {\bf 0.5539} \\
& ARIMA \cite{ariyo2014stock,box2015time} & 31.1630  & 117.0928 & 0.2091 & 0.9603 & 0.5590 & 0.4388 & 10.1357 & 10.6687 & 0.5230 & 0.3305 \\
& Prophet \cite{taylor2018forecasting}  & 33.0980 & 119.3963 & 0.3899 & 1.0763 & 0.5095 & 0.3780 & 9.9370 & 11.0289 & 0.5233 & 0.3203 \\
& DeepAR \cite{salinas2020deepar}\textsuperscript{$\dagger$}  & 13.3720 & 76.1662 & 0.1347 & 0.8909 & 0.6212 & 0.5091 & 9.3044 & 9.8906 & 0.6300 & 0.4095 \\
& LSTM \cite{hochreiter1997long}\textsuperscript{$\dagger$}  & 26.9251 & 88.0435 & 0.1894 & 0.9504 & 0.5721 & 0.4544 & 11.0396\textsuperscript{*} & 10.4210\textsuperscript{*} & 0.4983\textsuperscript{*} & 0.3407\textsuperscript{*} \\
& Informer \cite{zhou2021informer}\textsuperscript{$\dagger$}  & 12.7482 & 84.2029 & 0.0958 & 0.8630 & 0.7448 & 0.5207 & 9.5108\textsuperscript{*} & 10.2366\textsuperscript{*} & 0.6238\textsuperscript{*} & 0.4009\textsuperscript{*} \\
\toprule
\end{tabular}
\begin{flushleft}
\vspace{-0.12cm}
\scriptsize
\ * Since probability distributions can not be directly inferred from LSTM or Informer models, we use dropout during both training and inference to get 10 samples for each step.

\ $\dagger$ To enable a fair comparison, the information on sector and customer focus is one-hot encoded and concatenated with the decoder output.
\end{flushleft}
\vspace{0.3pt}
\caption{\label{table:quantitative}
The comparison with the popular and state-of-the-art baselines on two datasets with scaleup revenue time-series.}
\end{table*}

\subsection{Confidence Estimation}
\label{sec:algo_and_complexity}
We summarize the procedure of generating $T'$ revenues for company $c'$ in Algorithm~\ref{algo:inner_loop}, 
which outputs a $T'$-dimensional vector $[x_{T_{c'}+1}, x_{T_{c'}+2}, \ldots, x_{T_{c'}+T'}]$ representing one possible revenue trajectory. 
Because of the stochasticity of our prediction procedure, we might get a slightly different revenue trajectory if we run the same algorithm again.
As illustrated in Figure~\ref{fig:revenue_distribution}, this difference is insignificant initially (i.e. for $x_{T_{c'}+1}$), but it gets accumulated and amplified along with the extrapolation roll-out.
Intuitively, the uncertainty of the predicted revenue is reflected in these differences.
In order to capture that uncertainty in the predicted revenue trajectory, we choose to carry out $M\geq 10$ trials of revenue forecasting (i.e. Algorithm~\ref{algo:inner_loop}) with random seeds. 
It produces $M$ extrapolated revenue trajectories (for scaleup company $c'$), organized in a matrix:
\begin{equation}
\label{eq:multiple_predicted_trajectories}
  \widehat{\mathbf{X}}^{(c')}=
  \begin{bmatrix}
    x_{T_{c'}+1}^{\langle 1 \rangle} & x_{T_{c'}+2}^{\langle 1 \rangle} & \ldots & x_{T_{c'}+T'}^{\langle 1 \rangle} \\
    x_{T_{c'}+1}^{\langle 2 \rangle} & x_{T_{c'}+2}^{\langle 2 \rangle} & \ldots & x_{T_{c'}+T'}^{\langle 2 \rangle} \\
    \vdots & \vdots & \ddots & \vdots \\
    x_{T_{c'}+1}^{\langle M \rangle} & x_{T_{c'}+2}^{\langle M \rangle} & \ldots & x_{T_{c'}+T'}^{\langle M \rangle} \\
  \end{bmatrix}_{M\times T'},
\end{equation}
where the $m$-th row $\widehat{\mathbf{X}}^{(c')}_{\langle m, \cdot \rangle}\!=\!(x_{T_{c'}+1}^{\langle m \rangle}, x_{T_{c'}+2}^{\langle m \rangle}, \ldots, x_{T_{c'}+T'}^{\langle m \rangle})$ represents the $m$-th predicted revenue trajectory, and the $t'$-th column $\small\smash{\widehat{\mathbf{X}}^{(c')}_{\langle \cdot, t' \rangle}}$= $\small\smash{[x_{T_{c'}+t'}^{\langle 1 \rangle}, x_{T_{c'}+t'}^{\langle 2 \rangle}, \ldots, x_{T_{c'}+t'}^{\langle M \rangle}]^{\top}}$ contains a collection of predicted revenues for time point $T_{c'}\!+\!t'$ representing a revenue distribution for that period, as illustrated in Figure~\ref{fig:revenue_distribution}. 
The expected revenue at $T_{c'}+t'$ is the mean of $\small\smash{\widehat{\mathbf{X}}^{(c')}_{\langle \cdot, t' \rangle}}$, noted as $\bar{x}_{T_{c'}+t'}$. 
To capture uncertainty, we calculate the 95\% confidence interval (CI) $\bar{x}_{T_{c'}+t'}\pm\beta_{T_{c'}+t'}$ over $\small\smash{\widehat{\mathbf{X}}^{(c')}_{\langle \cdot, t' \rangle}}$ assuming a normal distribution:
\begin{equation}
\label{eq:confidence_interval_beta}
  \bar{x}_{T_{c'}+t'}\pm\beta_{T_{c'}+t'}=
  \bar{x}_{T_{c'}+t'}\pm\zeta\cdot\sqrt{\frac{\sum_{m=1}^M\left(x_{T_{c'}+t'}^{\langle m \rangle}-\bar{x}_{T_{c'}+t'}\right)^2}{M(M-1)}},
\end{equation}
where $\beta_{T_{c'}+t'}$ is the margin of error for the revenue prediction at time $T_{c'}+t'$. The constant $\zeta\approx1.96$ is the Z value corresponding to a 95\% confidence interval. Similarly, we should set $\zeta\approx1.645$ if we look for the 90\% confidence interval. 
Therefore, the prediction for the $c'$-th company $\widehat{\mathbf{x}}^{(c')}$ is noted as $\hat{\mathfrak{x}}^{(c')}$ in the form of \eqref{eq:revenue_prediction_company_c_prime}. 
Algorithm~\ref{algo:outer_loop} in the Appendix contains the pseudo code to produce the final resulting vector $\hat{\mathfrak{x}}^{(c')}$.

\subsection{Complexity and Limitation}

Algorithm~\ref{algo:inner_loop} is the pseudo code to output a $T'$-dimensional vector $[x_{T_{c'}+1}, x_{T_{c'}+2}, \ldots, x_{T_{c'}+T'}]$, formalized in Equation~\eqref{eq:final_revenue_forecasts}, representing one possible revenue trajectory.
The computational complexity of Algorithm~\ref{algo:inner_loop} is approximately $\mathcal{O}(C\times T_c^2)$. 
Algorithm~\ref{algo:outer_loop} describes the logic to produce the final resulting vector $\hat{\mathfrak{x}}^{(c')}$ formalized in Equation~\eqref{eq:revenue_prediction_company_c_prime}. 
To generate the confidence estimates, Algorithm~\ref{algo:outer_loop} runs Algorithm~\ref{algo:inner_loop} $M$ times, followed by an $\mathcal{O}(T')$ confidence estimation step. 
Since $M$ is a constant and $\mathcal{O}(T')$ has a lower order than $\mathcal{O}(C\times T_c^2)$, the overall computational complexity of SiRE is still $\mathcal{O}(C\times T_c^2)$. 
Figure \ref{fig:overall} depicts the overall SiRE approach outlining Algorithm~\ref{algo:inner_loop} and \ref{algo:outer_loop}, where it only requires revenue and sector information, thus fulfilling the 6th practical requirement in Section~\ref{sec:intro}.

The standards (e.g. IFRS and GAAP)\footnote{IFRS (international financial reporting standards) and GAAP (generally accepted accounting principles) are standards in EU and US, respectively. 
The most notable difference between them lies in the treatment of inventory, affecting revenue accounting.} that govern financial reporting and accounting vary from country to country.
SiRE implicitly assumes consistency of accounting standards, because how revenue is recognized is likely going to contribute to the applicability of SiRE.
To that end, we recommend readers to ensure the consistency of the accounting standard adopted when creating the dataset.
Finally, the authors would also like to point out that recurring revenue is not the only way to derive valuations for scaleups, rather a key aspect for many business models such as a subscription based one.

\section{Experiments}
The revenue time-series of scaleups are usually considered sensitive and typically not shared externally. Therefore, {\bf there has not been any public dataset available for benchmarking SiRE}. 
We have the advantage to gain access to such data from 
EQT Group\footnote{EQT is a global investment firm: {\color{blue}\url{https://eqtgroup.com}}} and 
Standard \& Poor's (S\&P) Capital IQ\footnote{A financial data provider: {\color{blue}\url{https://www.capitaliq.com}}}, forming two multi-sector datasets ARR129 and SapiQ respectively. 
ARR129 contains 1,485 monthly ARR (Annual Recurring Revenue) data points from 129 SaaS (Software as a Service) companies in growth stage. SapiQ comprises 766 yearly revenues from 158 scaleups headquartered in North America or Europe. 
ARR129 and SapiQ represent the common kind of revenue data possessed by PC firms. 
For parameter $r$ in Equations~\eqref{eq:dataset_measuring_subset} and \eqref{eq:forecast_measuring_dataset}, we search $\{\frac{1}{4}, \frac{1}{2}, \frac{3}{4}, 1\}$ and empirically discover that $r=\frac{1}{2}$ turns out to be a balanced choice. 
In confidence estimation (Section~\ref{sec:algo_and_complexity}), we use of $M=10$ for fast inference.
Because the scale of the datasets are small, we carry out the experiments on a single machine: Intel Quad-Core i7, 1.7-GHz CPU with 16-GB, 2133-MHz RAM.

\subsection{Quantitative Benchmarking}
We compare SiRE with five baseline methods that are either state-of-the-art in time-series prediction or widely adopted in revenue forecasting: ARIMA \cite{ariyo2014stock,box2015time}, Prophet \cite{taylor2018forecasting}, LSTM \cite{hochreiter1997long}, DeepAR \cite{salinas2020deepar}, and Informer \cite{zhou2021informer}. 
During evaluation, it is difficult to use classical cross validation because the observations are scarce and not exchangeable. We largely follow \cite{tashman2000out,taylor2018forecasting} to perform a ``rolling origin'' evaluation procedure by producing one extrapolation trajectory at each predefined cutoff point along the timeline. 
Since the feasible cutoff points and the length of the extrapolated trajectory can vary among different methods, we only include the common cutoff points and trajectories in metrics calculations. 
Since ARR129 and SapiQ are monthly and yearly datasets, we extrapolate 36 months and 12 years respectively. For yearly data, Equation~\eqref{eq:final_measurement} becomes $\small\smash{y_{t'+1}^{(c')} \coloneqq u_{t'}^{(c')}\!\!\cdot\!\hat{z}}$.
For LSTM and Informer, a confidence estimate is not available, hence we apply dropout during both training and inference to get 10 samples for each extrapolated step to approximate probability distributions.
Dropout can be viewed as a form of ensemble and regularization, which might have contradicting effect: 
ensemble boosts the performance while regularization might make the convergence harder on small datasets.   
We adopt three metrics (cf.~\cite{taylor2018forecasting,salinas2020deepar,ekambaram2020attention}) computed using the mean prediction: 
\begin{itemize}
\item root-mean-square error RMSE= $\small\smash{[\frac{1}{n}\sum_{i=1}^{n}(x - u)^2]^{1/2}}$,
\item mean absolute percentage error MAPE=$\small\smash{\frac{1}{n}\sum_{i=1}^{n}\left\vert\frac{u - x}{u}\right\vert}$,
\item Pearson correlation coefficient (PCC) that measures whether the actual ($u$) and predicted ($x$) time-series move in the same (+1) or opposite direction (-1).
\end{itemize}
We also compute two metrics (cf.~\cite{zhang2020large}) using the prediction distribution (the confidence is estimated from this distribution): 
\begin{itemize}
\item negative log-likelihood $\text{NLL} = -\frac{1}{n}\sum_{i=1}^{n}\log \hat{P}_\theta(u\vert x)$, where $\hat{P}_\theta$ is the probability density function.
\item 95\% CI accuracy (ACC) indicating if the booked revenue $u$ lies in the predicted 95\% CI.
\end{itemize}

In Table~\ref{table:quantitative}, the best results are emphasized in bold format. On both datasets, SiRE performs better than all other methods by a large margin over all five metrics. 
Despite of the simplicity and popularity of statistical methods (ARIMA and Prophet), they suffered severely from ignoring the time-series of other scaleups in similar sectors and stages, obtaining the worst result.
The RNN based approaches (LSTM and DeepAR) are greatly relieved from that problem by learning from all time-series up till the cutoff date, coinciding the findings in \cite{siami2018comparison}.
The Transformer based method, Informer, achieves the second best performance of all probably due to its effective attention mechanism. These experiments confirm that SiRE meets the 1st and 2nd practical requirements mentioned in Section \ref{sec:intro}.

\begin{figure}[t]
\centering
\subfigure{\includegraphics[height=0.8cm]{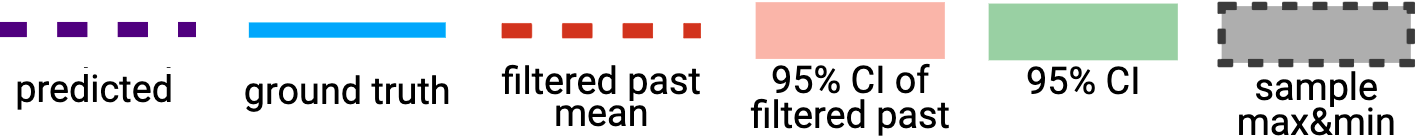}}\label{fig:vis_legend}\par%
\vspace{-0.15cm}
\setcounter{subfigure}{0}%
\subfigure[SiRE {\footnotesize(ours)}]{\includegraphics[height=2.92cm]{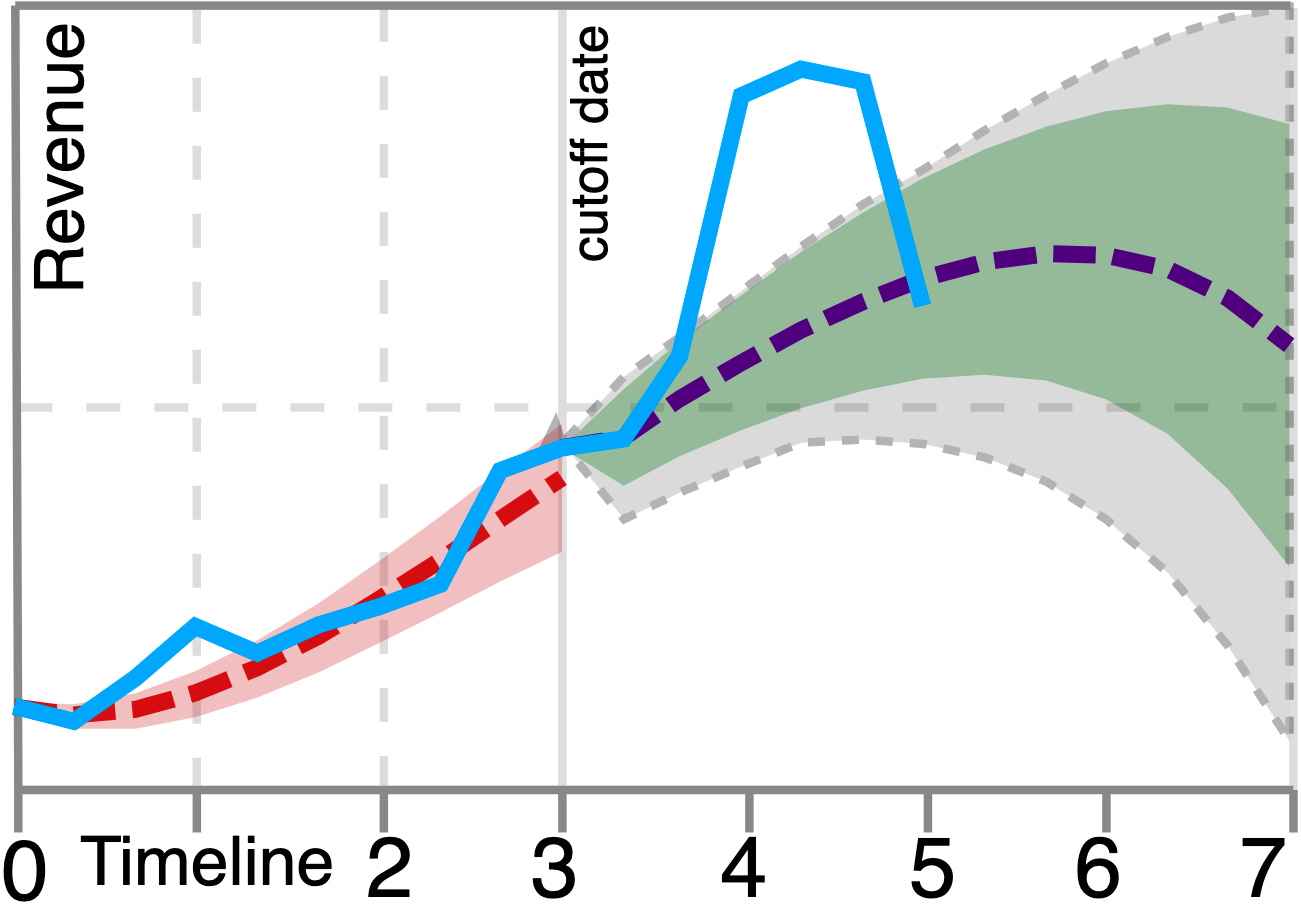}}\label{fig:vis_sire}
\subfigure[S][ARIMA {\footnotesize(statistical)}]{\includegraphics[height=2.92cm]{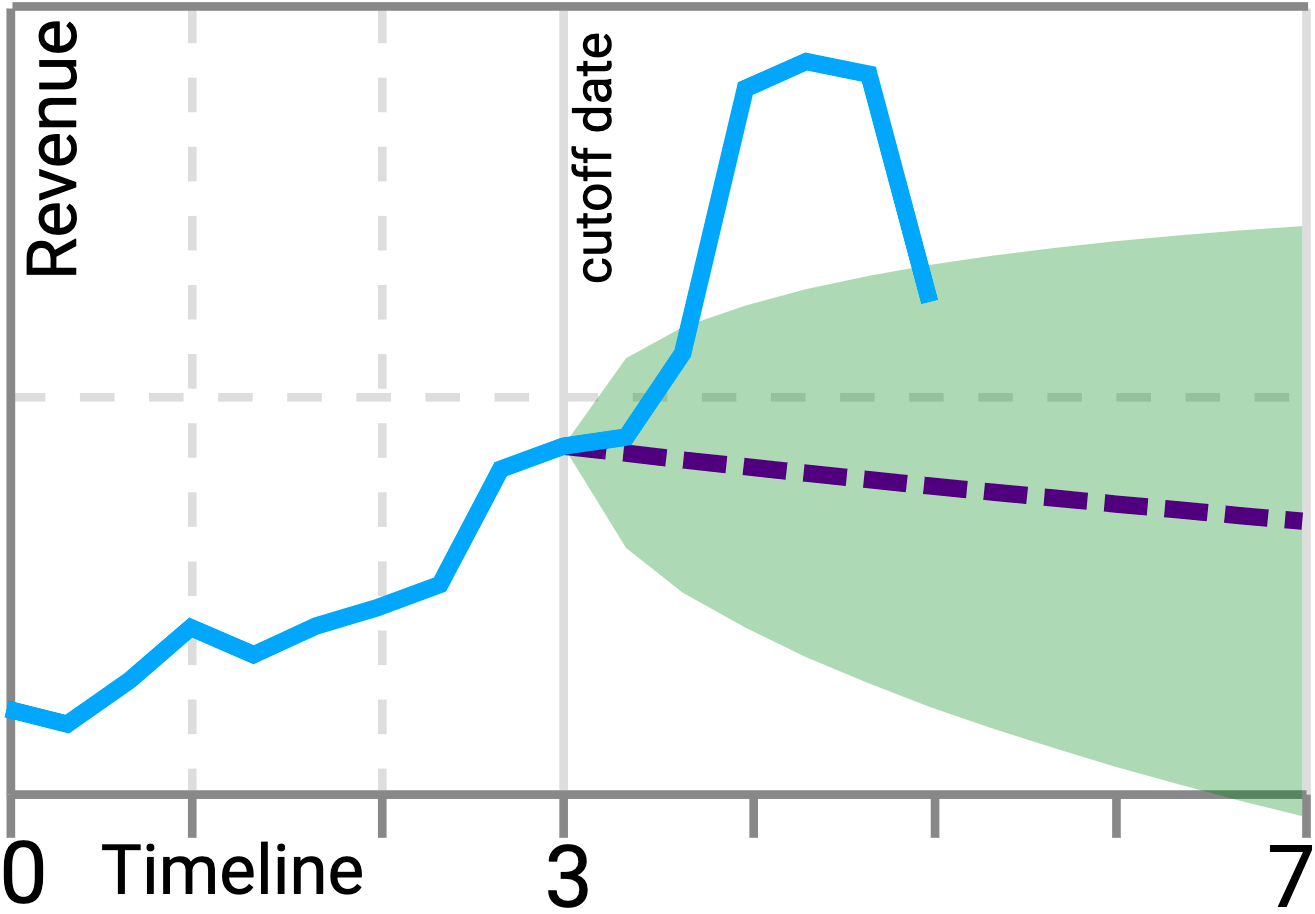}}\label{fig:vis_arima}\par%
\vspace{-0.1cm}
\subfigure[DeepAR {\footnotesize(RNN based)}]{\includegraphics[height=3.14cm]{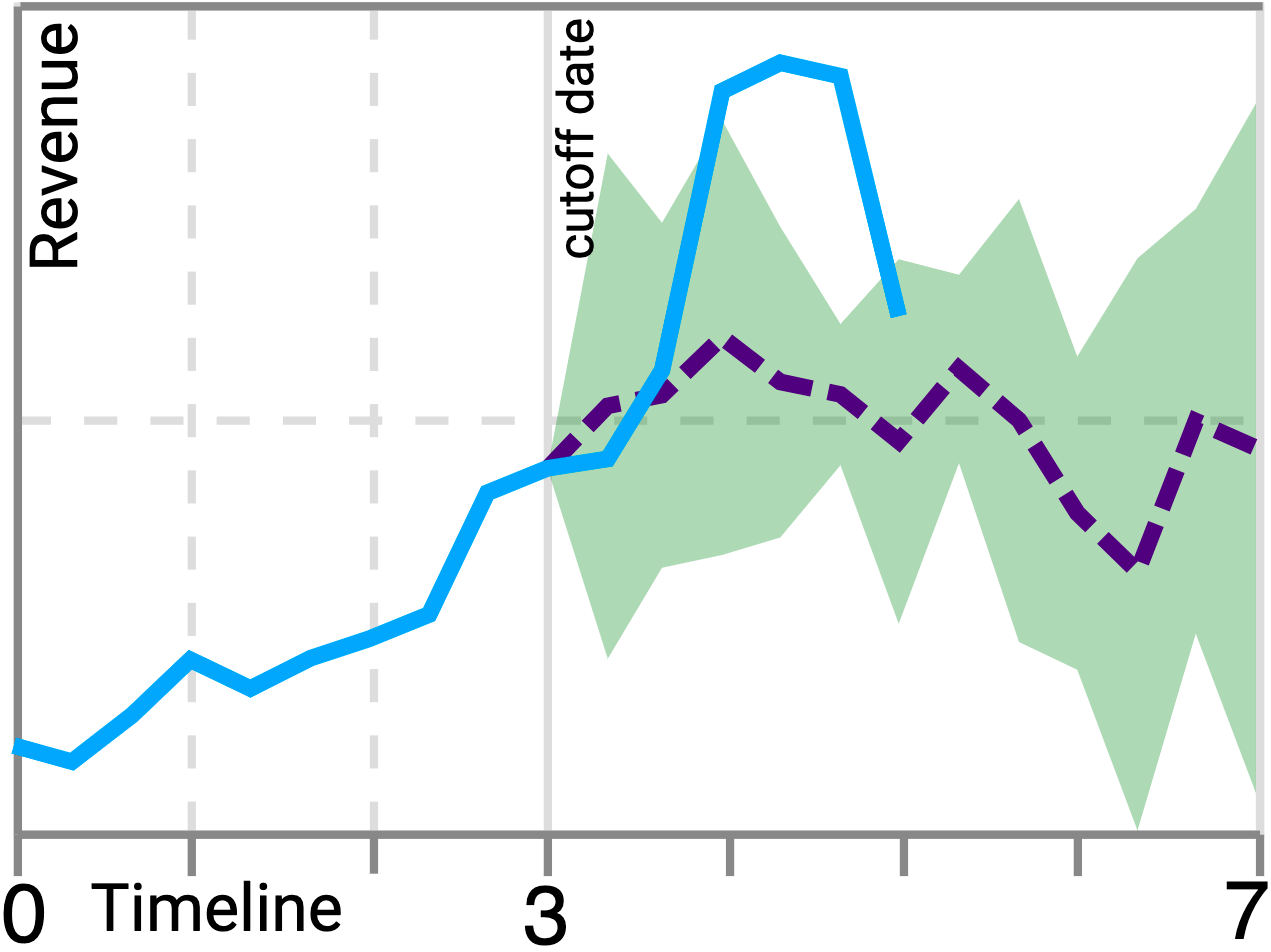}}\label{fig:vis_deepar}
\subfigure[Informer {\footnotesize (Transformer based)}]{\includegraphics[height=3.14cm]{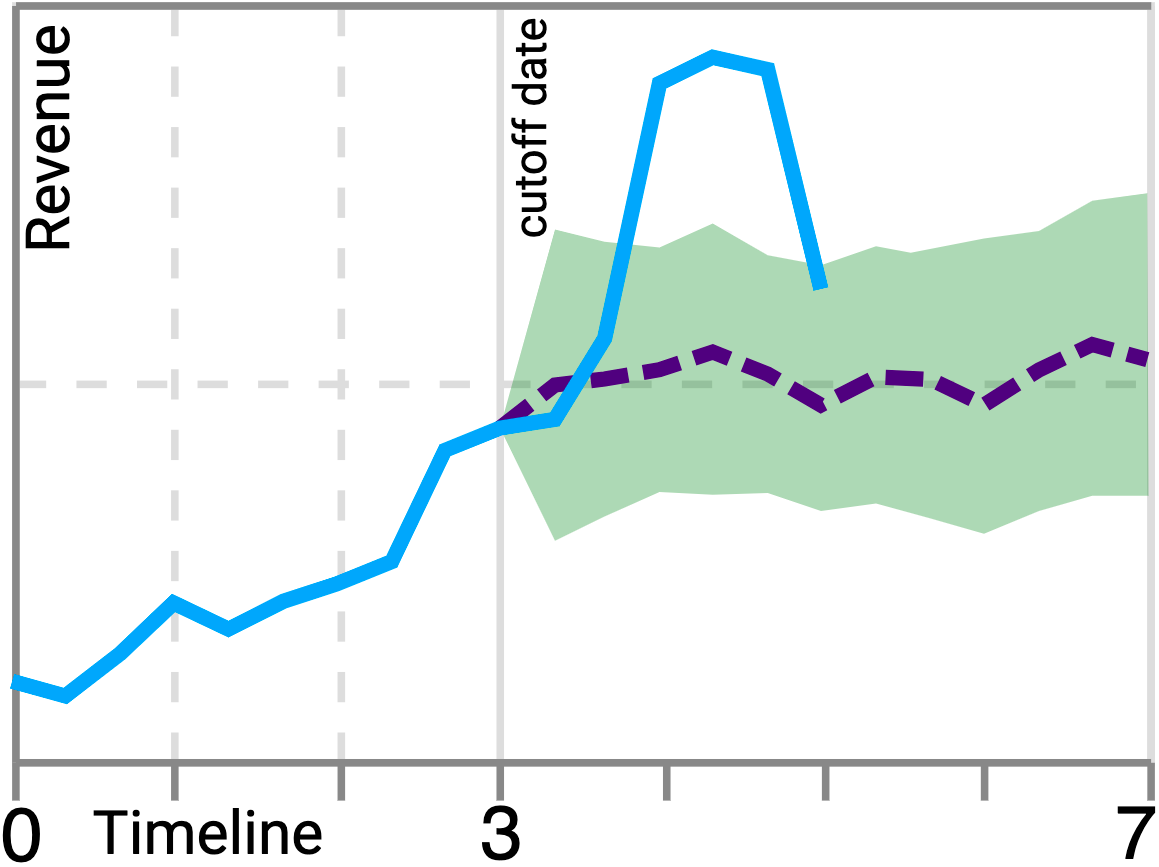}}\label{fig:vis_informer}
\begin{flushleft}
\vspace{-0.12cm}
\scriptsize
Note: The actual name and revenue (Y-axis) of the scaleup is regarded as sensitive information and therefore removed from the plots. The forecast starts from the ``cutoff date''.
\end{flushleft}
\vspace{-0.2cm}
\caption{The extrapolated revenue trajectories (from representative methods) for a scaleup from the SapiQ dataset.} \label{fig:vis}
\end{figure}

\subsection{Qualitative Inspection}
In Figure~\ref{fig:vis}, we visualize the extrapolated revenue trajectories for the same scaleup as a qualitative example. 
Besides SiRE, the inspected methods are ARIMA, DeepAR and Informer, which represent statistical, RNN-based and Transformer-based methodologies respectively. 
As a result, we will mainly report results from those methods in the rest of this paper.
Seen from the predicted mean (dashed purple line), ARIMA seems to be the least informative while DeepAR is the most volatile. 
SiRE turns out to be more advantageous than other baselines in several ways: 
(1) the missing historical data points are imputed; 
(2) the entire trajectory is smooth so that patterns and trends are easier to identify;
(3) the confidence can be naturally estimated everywhere, fulfilling the 5th practical requirement in Section \ref{sec:intro}; 
(4) the 95\% CI starts much narrower and grows much slower than the compared methods.

\subsection{Performance on Short Time-Series}
According to \eqref{eq:init_mu}, SiRE can start generating reasonable predictions using as few as three data points. 
The ability of obtaining reasonable revenue forecasts from short time-series is valued by investment professionals. 
In Table~\ref{table:short_ts}, we measure the quality of extrapolations based on time-series with only three (e.g. ``RMSE@3m'') and six (e.g. ``RMSE@6m'') data points.
Because ARR129 contains monthly revenue data that has an average span of 12 months (i.e. 12 data points), ARR129 is better suited for this evaluation.
Table~\ref{table:short_ts} shows that SiRE performs the best (indicated in boldface) in most of the cases, thereby satisfying the 3rd practical requirement in Section \ref{sec:intro}.
Compared to the overall results in Table~\ref{table:quantitative}, all methods suffered greatly when dealing with short time-series, and their performance generally improves when the length of the time-series increases. 

\begin{table}[t]
\small
\addtolength{\tabcolsep}{2.3pt}
\renewcommand{\arraystretch}{0.9}
\centering
\begin{tabular}{c|r|r|r|r|r}
\bottomrule
\multicolumn{2}{c|}{\textbf{Methods}:} & \multicolumn{1}{c|}{SiRE} & \multicolumn{1}{c|}{ARIMA} & \multicolumn{1}{c|}{DeepAR} & \multicolumn{1}{c}{Informer} \\
\hline
\parbox[t]{2mm}{\multirow{10}{*}{\rotatebox[origin=c]{90}{Metrics (ARR129)}}} 
& RMSE@3m & {\bf 18.3391} & 40.0290 & 19.4071 & 27.7033 \\
& MAPE@3m & {\bf 0.1004}  & 0.2960 & 0.2005 & 0.2638 \\
& PCC@3m & {\bf 0.7038}  & 0.5602 & 0.6084 & 0.5804 \\
& NLL@3m & 9.9370  & {\bf 9.9016} & 10.6430 & 10.6975\textsuperscript{*} \\
& ACC@3m & 0.5609  & {\bf 0.5704} & 0.5619 & 0.5492\textsuperscript{*} \\
\cline{2-6} 
& RMSE@6m & {\bf 14.0432}  & 37.0230 & 18.3806 & 17.2713 \\
& MAPE@6m & {\bf 0.0831}  & 0.2837 & 0.1897 & 0.1538 \\
& PCC@6m & {\bf 0.7540}  & 0.5486 & 0.6291 & 0.6492 \\
& NLL@6m & {\bf 9.1903}  & 9.6160 & 10.3031 & 11.9310\textsuperscript{*} \\
& ACC@6m & {\bf 0.6315}  & 0.5602 & 0.6114 & 0.5872\textsuperscript{*} \\
\toprule
\end{tabular}
\begin{flushleft}
\vspace{-0.12cm}
\scriptsize
\ * Dropout is applied in training and inference to get 10 predicted revenues for each month.
\end{flushleft}
\vspace{1.5pt}
\caption{The performance of revenue extrapolation from short time-series in the ARR129 dataset: ``@3m'' and ``@6m'' stands for extrapolation using only 3 and 6 data points (i.e. months), respectively. The best results are in bold.}
\label{table:short_ts}
\vspace{-0.3cm}
\end{table}

\subsection{Performance on Long-Term Forecast}
Accurate long-term revenue forecast is one of the key requirements when evaluating a scaleup prior to making an investment decision. 
To benchmark SiRE's performance on predicting long-term revenue, we calculate five metrics (cf.~Table~\ref{table:quantitative} and \ref{table:short_ts}) using the predicted revenues for either the 2nd-to-3rd (e.g. ``RMSE$\rhd$2-3y'') or the 4th-to-5th (e.g. ``RMSE$\rhd$4-5y'') year in the future. 
As shown in Table~\ref{table:long-term}, we report the results on the SapiQ dataset, where each scaleup has five annual revenue points on average. 
Because it is not viable to calculate PCC using only one or two predicted samples, we exclude PCC from the comparison.
Generally speaking, the further into the future, the more challenging the prediction becomes. 
Nonetheless, SiRE achieves the best performance (in both cases of ``$\rhd$2-3y'' and ``$\rhd$4-5y'') when predicting long-term revenue. This fulfills the 4th practical requirement in Section \ref{sec:intro}.

\begin{table}[t]
\small
\addtolength{\tabcolsep}{1.9pt}
\renewcommand{\arraystretch}{0.9}
\centering
\begin{tabular}{c|r|r|r|r|r}
\bottomrule
\multicolumn{2}{c|}{\textbf{Methods}:} & \multicolumn{1}{c|}{SiRE} & \multicolumn{1}{c|}{ARIMA} & \multicolumn{1}{c|}{DeepAR} & \multicolumn{1}{c}{Informer} \\
\hline
\parbox[t]{2mm}{\multirow{8}{*}{\rotatebox[origin=c]{90}{Metrics (SapiQ)}}} 
& RMSE$\rhd$2-3y & {\bf 52.7403} & 110.0589 & 75.4020 & 78.0230 \\
& MAPE$\rhd$2-3y & {\bf 0.6038}  & 0.9562 & 0.8940 & 0.8373 \\
& NLL$\rhd$2-3y & {\bf 8.3057}  & 10.0433 & 9.7248 & 10.5820\textsuperscript{*} \\
& ACC$\rhd$2-3y & {\bf 0.5730}  & 0.3200 & 0.4327 & 0.4105\textsuperscript{*} \\
\cline{2-6} 
& RMSE$\rhd$4-5y & {\bf 79.2835}  & 123.6132 & 88.9603 & 86.6090 \\
& MAPE$\rhd$4-5y & {\bf 0.9034}  & 1.1250 & 1.0176 & 0.9639 \\
& NLL$\rhd$4-5y & {\bf 9.6046}  & 12.2188 & 11.2290 & 11.4850\textsuperscript{*} \\
& ACC$\rhd$4-5y & {\bf 0.4284}  & 0.3554 & 0.4046 & 0.3991\textsuperscript{*} \\
\toprule
\end{tabular}
\begin{flushleft}
\vspace{-0.12cm}
\scriptsize
\ * Dropout is applied in training and inference to get 10 predicted revenues for each year.
\end{flushleft}
\vspace{1.5pt}
\caption{The performance of long-term revenue forecast on the SapiQ dataset: ``$\rhd$2-3y'' and ``$\rhd$4-5y'' stands for revenue forecast for the 2nd-to-3rd and the 4th-to-5th year in the future, respectively. The best results are in bold.}
\label{table:long-term}
\vspace{-0.2cm}
\end{table}

\subsection{On Training and Inference Efficiency}
Practically, the revenue dataset $\mathbf{U}$ [cf. Equation \eqref{eq:dataset_U}] changes frequently because of (1) incoming scaleup companies, (2) new revenue data points for existing scaleups, and (3) removal of incorrect or outdated data.
The model hence needs to adapt to any data change timely and efficiently, which fulfills the 7th practical requirement mentioned in Section~\ref{sec:intro}.
To evaluate this, we measure the training and inference time for SiRE, ARIMA, DeepAR and Informer on both datasets. 
Table~\ref{table:train-inference-time} shows that SiRE is largely on par with statistical methods like ARIMA, and more efficient than DL-based models, by a large margin.
Specifically, SiRE and ARIMA can incorporate data change every second while other models need at least 5 minutes.
Moreover, the superior performance of SiRE compared to ARIMA (cf. Tables~\ref{table:quantitative}-\ref{table:long-term}) makes SiRE the best methodological choice for the benchmarking datasets.

\begin{table}[t]
\small
\addtolength{\tabcolsep}{0.1pt}
\renewcommand{\arraystretch}{0.9}
\centering
\begin{tabular}{c|l|r r r r}
\bottomrule
\multicolumn{2}{c|}{\textbf{Methods}:} & \multicolumn{1}{c}{SiRE\textsuperscript{*}} & \multicolumn{1}{c}{ARIMA\textsuperscript{$\dagger$}} & \multicolumn{1}{c}{DeepAR} & \multicolumn{1}{c}{Informer} \\
\hline
\multirow{2}{*}{\specialcell{ARR129 \\ (1,485)\textsuperscript{$\ddagger$}}}
& Training (sec.) & 1.180 & 0.896 & 294.460 & 467.412 \\
& Inference (sec.) & 1.206  & 0.763 & 1.139 & 1.439 \\
\cline{1-6} 
\multirow{2}{*}{\specialcell{SapiQ \\ (766)\textsuperscript{$\ddagger$}}}
& Training (sec.) & 0.958  & 0.804 & 243.592 & 420.340 \\
& Inference (sec.) & 1.054  & 0.682 & 1.094 & 1.277 \\

\toprule
\end{tabular}
\begin{flushleft}
\vspace{-0.12cm}
\scriptsize
\ * The current implementation of SiRE does not strive for high performance.

\ $\dagger$ ARIMA training does not require traversing other scaleup companies.

\ $\ddagger$ The number of revenue data points in each dataset.
\end{flushleft}
\vspace{2pt}
\caption{The average training and inference time (measured in seconds) required by different methods on two datasets.}
\label{table:train-inference-time}
\end{table}

\subsection{On Explainability of SiRE Forecasts}
\begin{figure}[t]
  \centering
  \includegraphics[width=\linewidth]{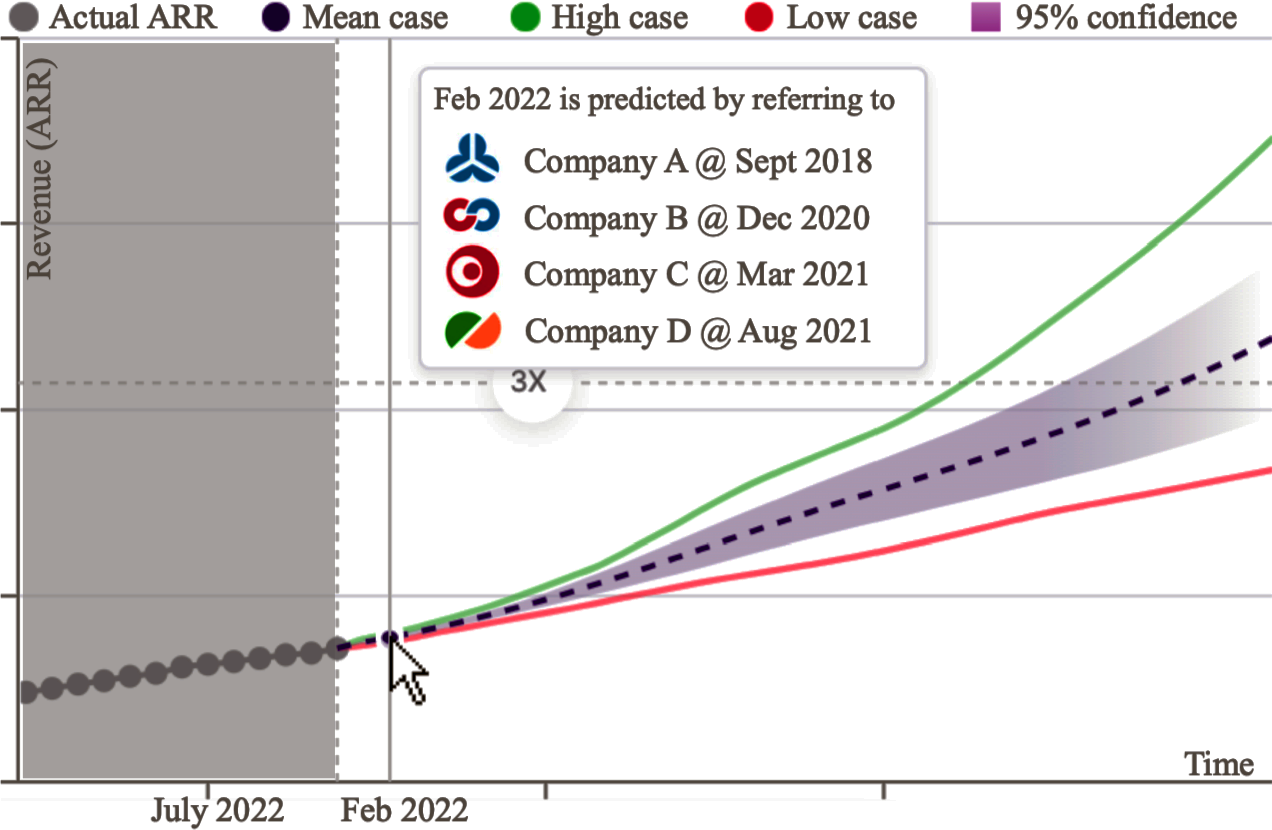}
  \caption{Explaining the extrapolated revenue from SiRE: a screenshot of our investment platform (EQT Motherbrain).}
\label{fig:app}
\end{figure}

When predicting each revenue point, our measuring approach puts together a measuring dataset $\overline{\mathbf{U}}'$ [cf. Equation~\eqref{eq:forecast_growth_subset}] containing fragments of revenue time-series from other scaleups, which can provide direct explainability to SiRE results. 
Figure~\ref{fig:app} demonstrates how one can interactively interpret the extrapolated revenue via our investment platform (Motherbrain)\footnote{{\color{blue}\url{https://eqtgroup.com/motherbrain}} and {\color{blue}\url{https://eqtventures.com/motherbrain}}} with SiRE integrated: when the user's mouse cursor hovers over a certain predicted point, some companies from $\overline{\mathbf{U}}'$ (with a past calendar date) are shown as the similar ones to the scaleup-in-scope at the date of prediction. 
In the example demonstrated in Figure~\ref{fig:app}, company A, B, C and D (at particular points in time) contribute the most to the predicted revenue on February 2022. 
This information provides the investment professionals (the users) with insight into how the prediction was generated, which has two major benefits: (1) it can increase acceptance of the prediction by providing transparency; (2) it can result in valuable feedback to improve the algorithm and/or the definition of $\overline{\mathbf{U}}'$. This satisfies the 8th practical requirement mentioned in Section \ref{sec:intro}. 

\subsection{Evaluation from the perspective of investment professionals}
From the perspective of PC investment professionals, it is not critical to have each extrapolated revenue point close to the ground truth. 
To make an informed investment decision, it is of great importance to understand if the revenue will change significantly several years down the line.
For instance, if the investment target demands that the revenue should be at least 3 times the value at the time of investment (so called 3x) after 4 or 5 years, then the model should be able to identify such scaleup companies (noted as ``>3x\_in\_4/5y'') as accurately as possible.
This is equivalent to computing the true positive rate (TPR) of ``>3x\_in\_4/5y'' cases.
We test several such cases on both datasets, and present the results in Table~\ref{table:investors-view}.
Although SiRE displays the best performance from the investors' point of view,
its TPR scores struggle to surpass 60\% for SapiQ and 70\% for ARR129.
Those results might improve when incorporating more context features such as country and size of the company.
\begin{table}[t]
\small
\addtolength{\tabcolsep}{-0.4pt}
\renewcommand{\arraystretch}{0.9}
\centering
\begin{tabular}{c|l|r r r r}
\bottomrule
\multicolumn{2}{c|}{\textbf{Methods}:} & \multicolumn{1}{c}{SiRE} & \multicolumn{1}{c}{ARIMA} & \multicolumn{1}{c}{DeepAR} & \multicolumn{1}{c}{Informer} \\
\hline
\multirow{2}{*}{ARR129}
& TPR>2x\_in\_2/3y & {\bf 0.6938} & 0.5560 & 0.5904 & 0.5875 \\
& TPR>3x\_in\_2/3y & {\bf 0.6809} & 0.5100 & 0.6060 & 0.5723 \\
\cline{1-6} 
\multirow{4}{*}{SapiQ}
& TPR>2x\_in\_2/3y & {\bf 0.5780} & 0.3903 & 0.4316 & 0.4126 \\
& TPR>3x\_in\_2/3y & {\bf 0.5402} & 0.3729 & 0.4040 & 0.3890 \\
& TPR>3x\_in\_4/5y & {\bf 0.5537} & 0.3505 & 0.3830 & 0.4057 \\
& TPR>4x\_in\_4/5y & {\bf 0.5290} & 0.3511 & 0.3714 & 0.3965 \\
\toprule
\end{tabular}

{\scriptsize E.g. TPR>2x\_in\_2/3y is the true positive rate of scaleups that reach 2x revenue after 2 or 3 years.}
\vspace{3.5pt}
\caption{The evaluation results from the perspective of investment professionals.}
\label{table:investors-view}
\vspace{-0.8cm}
\end{table}

\section{Conclusion}
In this work, we propose the SiRE algorithm to unlock the possibility of automating long-term fine-grained revenue extrapolation using scarce data.
SiRE models the revenue dynamics as a specific LDS which is solved using the EM algorithm, where the core innovation lies in how we choose to obtain noisy revenue measurements.
By design, it works for scaleups that operate in various sectors and provides confidence estimates.
The quantitative experiments on two datasets show that SiRE significantly surpasses the popular and state-of-the-art baselines on all carefully selected metrics. 
The qualitative analysis illustrates some advantageous attributes of SiRE, such as imputation of missing data, smoothness of trajectories and explainability of predictions. We also observed great performance when SiRE extrapolates long-term predictions from short time-series.
SiRE is an agile algorithm that, due to its high training and inference efficiency, adapts effortlessly to data change.
For investment professionals, SiRE can precisely find scaleup companies that have great potential of a revenue uplift in 2 to 5 years.
Future work includes studies on filter ablation and applicability to other relevant metrics such as customer churn, life-time value and conversion rate.

\begin{acks}
To the EQT Motherbrain team, for the support and interest in this work.
To Xiaolong Liu (Intel Lab) and Wenbing Huang (Tsinghua University) for discussions around Kalman filter and LDS. 
\end{acks}

\bibliographystyle{ACM-Reference-Format}
\bibliography{sire-ref}

\pagebreak
\appendix

\section{Derive the Optimization Target}
\label{sec:derive_opt_target}
The revenue model unfolds recursively from the initial state $\mathbf{x}_1$, which is a normal random vector with mean vector $\boldsymbol{\mu}\in\mathbb{R}^5$ and a $5\times 5$ covariance matrix $\boldsymbol{\Omega}$. 
The optimizable parameters of the model are $\mathbf{Q}$, $\mathbf{R}$, $\boldsymbol{\mu}$, and $\boldsymbol{\Omega}$. Note that only the output (the measured revenue $\mathbf{y}_{t}$) of the system can be observed while the state and all the noise variables are hidden. 
Because there is no input to the system, solving \eqref{eq:lds_dynamics} can be seen as addressing an unsupervised problem. Specifically, the goal is to model the unconditional joint density of $\{y\}=(y_1,y_2,\ldots,y_T)$ and $\{\mathbf{x}\}=(x_1,x_1,\ldots,x_T)$:
\begin{equation}
\label{eq:unconditional_joint_density}
p(\{\mathbf{x}\},\{y\})=
  p(\mathbf{x}_1) \cdot
  \prod_{t=2}^T p(\mathbf{x}_t\vert \mathbf{x}_{t-1}) \cdot
  \prod_{t=1}^T p(y_t\vert \mathbf{x}_{t}).
\end{equation}
For the convenience of derivation, rather than regarding the state $\mathbf{x}_t$ as a deterministic value corrupted by random noise, we treat it as a single Gaussian random variable; based on \eqref{eq:lds_dynamics}, the conditional state density can be written as
\begin{equation}
\label{eq:conditional_state_density}
p(\mathbf{x}_t\vert \mathbf{x}_{t-1})=
(2\pi)^{\frac{5}{2}} \cdot
\vert \mathbf{Q} \vert^{-\frac{1}{2}} \cdot
e^{
  -\frac{1}{2}(\mathbf{x}_t-\mathbf{A}\mathbf{x}_{t-1})^\top
  \mathbf{Q}^{-1}
  (\mathbf{x}_t-\mathbf{A}\mathbf{x}_{t-1})
};
\end{equation}
we could form a similar combination for the measurements and obtain the conditional density for $y_t$:
\begin{equation}
\label{eq:conditional_measurement_density}
p(y_t\vert \mathbf{x}_t)=
(2\pi)^{\frac{1}{2}} \cdot
\vert \mathbf{R} \vert^{-\frac{1}{2}} \cdot
e^{
  -\frac{1}{2}(y_t-\mathbf{c}\mathbf{x}_{t})^2
  \mathbf{R}^{-1}
}.
\end{equation}
Assuming a Gaussian initial state density, we have
\begin{equation}
\label{eq:initial_state_density}
p(\mathbf{x}_1)=
(2\pi)^{\frac{5}{2}} \cdot
\vert \boldsymbol{\Omega} \vert^{-\frac{1}{2}} \cdot
e^{
  -\frac{1}{2}(\mathbf{x}_1-\boldsymbol{\mu})^\top
  \boldsymbol{\Omega}^{-1}
  (\mathbf{x}_1-\boldsymbol{\mu})
}.
\end{equation}
By solving equations~\eqref{eq:unconditional_joint_density} to \eqref{eq:initial_state_density}, the joint log likelihood of the complete data becomes
\begin{equation}
\label{eq:joint_log_likelihood}
\small
\begin{aligned}
  \log p(\{\mathbf{x}\},\{y\}) \mathring{=} & -\textstyle{\sum}_{t=1}^T \frac{1}{2}(y_t-\mathbf{c}\mathbf{x}_{t})^2
  \mathbf{R}^{-1} - \frac{T}{2}\log\vert \mathbf{R}\vert \\
  & -\textstyle{\sum}_{t=2}^T \frac{1}{2}(\mathbf{x}_t\!-\!\mathbf{A}\mathbf{x}_{t-1})^\top
  \mathbf{Q}^{-1}
  (\mathbf{x}_t\!-\!\mathbf{A}\mathbf{x}_{t-1})\!-\!\frac{T-1}{2}\log\vert \mathbf{Q}\vert \\
  & \textstyle - \frac{1}{2}(\mathbf{x}_1-\boldsymbol{\mu})^\top
  \boldsymbol{\Omega}^{-1}
  (\mathbf{x}_1-\boldsymbol{\mu}) - \frac{1}{2}\log\vert \boldsymbol{\Omega}\vert 
  \; ,
\end{aligned}
\end{equation}
which is to be maximized with respect to the parameters $\mathbf{Q}$, $\mathbf{R}$, $\boldsymbol{\mu}$, and $\boldsymbol{\Omega}$.
Since $\log p(\{\mathbf{x}\},\{y\})$ depends on the latent states $\{\mathbf{x}\}$ that are unobservable, we largely follow \cite{ghahramani1996parameter,shumway1982approach} to apply the EM algorithm \cite{dempster1977maximum} conditionally with respect to the measured revenue series $\{y\}$, that is to maximize the following expectation $\mathfrak{O}$:
\begin{equation}
\label{eq:conditional_log_likelihood_appendix}
\begin{aligned}
&\begin{aligned}
\mathfrak{O} = & \;\mathbb{E}[\log p(\{\mathbf{x}\},\{y\})\vert\{y\}] \\
  = & -\frac{1}{2}\Tr\left\{\mathbf{R}^{-1}\textstyle{\sum}_{t=1}^T \left[(y_t-\mathbf{c}\mathbf{x}_t^T)^2
  +\mathbf{c}\mathbf{P}_t^T\mathbf{c}^\top
   \right]\right\}
  - \frac{T}{2}\log\vert \mathbf{R}\vert \\
  & -\frac{1}{2}\Tr\left\{\mathbf{Q}^{-1}(\mathbf{G}-\mathbf{F}\mathbf{A}^{\top}\!-\!\mathbf{A}\mathbf{F}^{\top}\!+\mathbf{A}\mathbf{E}\mathbf{A}^{\top})\right\}
  - \frac{T\!-\!1}{2}\log\vert \mathbf{Q}\vert \\
  & - \frac{1}{2}\Tr\left\{\boldsymbol{\Omega}^{-1}\left[\mathbf{P}_1^T+(\mathbf{x}_1^T-\boldsymbol{\mu})(\mathbf{x}_1^T-\boldsymbol{\mu})^\top\right]\right\}
  - \frac{1}{2}\log\vert \boldsymbol{\Omega}\vert,
\end{aligned} \\
&\begin{aligned}
  \text{where }\mathbf{E}= & \textstyle{\sum}_{t=2}^T\left[\mathbf{P}_{t-1}^T+\mathbf{x}_{t-1}^T(\mathbf{x}_{t-1}^T)^\top\right], \\
  \mathbf{G}= & \textstyle{\sum}_{t=1}^T\left[\mathbf{P}_t^T+\mathbf{x}_t^T(\mathbf{x}_t^T)^\top\right]\; \text{and} \\
  \mathbf{F}= & \textstyle{\sum}_{t=2}^T\left[\mathbf{P}_{t,t-1}^T+\mathbf{x}_t^T(\mathbf{x}_{t-1}^T)^\top\right].
\end{aligned}
\end{aligned}
\end{equation}
The operation $\Tr(\cdot)$ denotes the trace calculation, $\smash{\{y\}_{t_0}^{t_1}}$ represents a sub-sequence of $\{y\}$, i.e. $\small\smash{\{y\}_{t_0}^{t_1}=(y_{t_0}, y_{t_0+1}, \ldots, y_{t_1})}$, and $\smash{\mathbf{x}_t^\tau}$ defines the conditional mean $\small\smash{\mathbb{E}(\mathbf{x}_t\vert \{y\}_1^\tau)}$. 
The terms $\smash{\mathbf{P}_t^{\tau}}$ and $\smash{\mathbf{P}_{t, t-1}^{\tau}}$ define the covariances $\smash{\Cov(\mathbf{x}_t\vert\{y\}_1^\tau)}$ and $\smash{\Cov(\mathbf{x}_t\mathbf{x}_{t-1}\vert\{y\}_1^\tau)}$, respectively.

\section{Expectation Maximization}
\label{sec:em-detail}
\subsection{E-step}
The expectation step embodies two consecutive sub-steps: forward filtering and backward smoothing. This is a process of obtaining $\mathbf{x}_t$ from  $y_t$ represented by the green arrows in Figure~\ref{fig:measuring_revenue}. The formulations of the E-step are mostly adapted from \cite{byron2004derivation}. 

\subsubsection{Forward filtering}
\label{sec:em-forward-filtering}
We first seek the values of $\smash{\mathbf{x}_t^t}$ and $\smash{\mathbf{P}_t^t}$ with initial settings $\smash{\mathbf{x}_1^0=\boldsymbol{\mu}}$ and $\smash{\mathbf{P}_1^0=\boldsymbol{\Omega}}$, which is usually called forward filtering due to the fact that the parameters are recursively calculated from the beginning of the time-series.
\begin{equation}
\label{eq:forward_filtering}
\begin{aligned}
  \mathbf{x}_t^{t-1} = & \mathbf{A}\mathbf{x}_{t-1}^{t-1}\;, \\
  \mathbf{P}_t^{t-1} = & \mathbf{A}\mathbf{x}_{t-1}^{t-1}\mathbf{A}^\top+\mathbf{Q}\;, \\
  \mathbf{K}_t = & \mathbf{P}_t^{t-1}\mathbf{c}^\top(\mathbf{c}\mathbf{P}_t^{t-1}\mathbf{c}^\top + \mathbf{R})^{-1}\;\;, \\
  \mathbf{x}_t^t = & \mathbf{x}_t^{t-1} + \mathbf{K}_t(y_t - \mathbf{c}\mathbf{x}_t^{t-1})\;, \\
  \mathbf{P}_t^t = & \mathbf{P}_t^{t-1} - \mathbf{K}_t\mathbf{c}\mathbf{P}_t^{t-1}\;.
\end{aligned}
\end{equation}

\subsubsection{Backward smoothing}
Now we seek the values for mean $\smash{\mathbf{x}_t^T}$ and variance $\smash{\mathbf{P}_t^T}$, which differs from the ones (i.e. $\smash{\mathbf{x}_t^t}$ and $\smash{\mathbf{P}_t^t}$) computed in \eqref{eq:forward_filtering} in that they depend on both past and future measurements. This calculation is called backward smoothing since it starts from the tail of the time-series conditioned by all available measurements.
\begin{equation}
\label{eq:backward_smoothing_1}
\begin{aligned}
  \mathbf{J}_{t-1} = & \mathbf{P}_{t-1}^{t-1}\mathbf{A}^\top(\mathbf{P}_t^{t-1})^{-1}\;, \\
  \mathbf{x}_{t-1}^T = & \mathbf{x}_{t-1}^{t-1} + \mathbf{J}_{t-1}(\mathbf{x}_t^T - \mathbf{A}\mathbf{x}_{t-1}^{t-1})\;, \\
  \mathbf{P}_{t-1}^T = & \mathbf{P}_{t-1}^{t-1} + \mathbf{J}_{t-1}(\mathbf{P}_t^T - \mathbf{P}_{t-1}^{t-1})\mathbf{J}_{t-1}^\top\;.
\end{aligned}
\end{equation}

In the upcoming M-step, we will also need to calculate the expectation matrices $\mathbb{E}(\mathbf{x}_t\mathbf{x}_t^\top\vert \{y\})$ and $\mathbb{E}(\mathbf{x}_t,\mathbf{x}_{t-1}^\top\vert \{y\})$ that are specifically denoted by 
$\mathbf{P}_t$ and $\mathbf{P}_{t, t-1}$, respectively. They can be obtained using the backward recursions:
\begin{equation}
\label{eq:backward_smoothing_2}
\begin{aligned}
  \mathbf{P}_t= &\mathbf{P}_t^T + \mathbf{x}_t^T(\mathbf{x}_t^T)^\top\;, \\
  \mathbf{P}_{T, T-1}^T = & (\mathbf{I} - \mathbf{K}_T\mathbf{c})\mathbf{A}\mathbf{P}_{T-1}^{T-1}\;, \\
  \mathbf{P}_{t-1, t-2}^T = & \mathbf{P}_{t-1}^{t-1}\mathbf{J}_{t-2}^\top + \mathbf{J}_{t-1}(\mathbf{P}_{t, t-1}^T - \mathbf{A}\mathbf{P}_{t-1}^{t-1})\mathbf{J}_{t-2}^\top\;, \\
  \mathbf{P}_{t, t-1}= &\mathbf{P}_{t, t-1}^T + \mathbf{x}_t^T(\mathbf{x}_{t-1}^T)^\top\;,
\end{aligned}
\end{equation}
where $\mathbf{I}$ is an identity matrix. The terms $\mathbf{K}_T$, $\mathbf{x}_t^T$, $\mathbf{P}_t^T$, $\mathbf{P}_{t-1}^{t-1}$, $\mathbf{P}_{T-1}^{T-1}$, $\mathbf{J}_{t-1}$, $\mathbf{J}_{t-2}$, and $\mathbf{x}_{t-1}^T$ have already been calculated in \eqref{eq:forward_filtering} and \eqref{eq:backward_smoothing_1}.

\subsection{M-step}
\label{sec:m-step}
In the M-steps, each free parameter (i.e. $\mathbf{Q}$, $\mathbf{R}$, $\boldsymbol{\mu}$, and $\boldsymbol{\Omega}$) is re-estimated by taking the corresponding partial derivative of \eqref{eq:conditional_log_likelihood}, setting it to zero, and solving. 
Results are taken from \cite{ghahramani1996parameter}:
\begin{itemize}
\item Measurement noise covariance $\mathbf{R}$: 
\begin{equation*}
\label{eq:new_R}
\small
\mathbf{R}^{(\text{new})} = -\frac{1}{T} \textstyle{\sum}_{t=1}^T\left(y_t^2-y_t\mathbf{c}\mathbf{x}_t^T\right).
\end{equation*}
\item Latent state noise covariance $\mathbf{Q}$:
\begin{equation*}
\label{eq:new_Q}
\small
\mathbf{Q}^{(\text{new})} = -\frac{1}{T-1} \left(\textstyle{\sum}_{t=2}^T\mathbf{P}_t - \mathbf{A}\textstyle{\sum}_{t=2}^T\mathbf{P}_{t-1, t}\right).
\end{equation*} 
\item The mean of initial latent state $\boldsymbol{\mu}^{(\text{new})} = \mathbf{x}_1^T$.
\item The covariance of initial latent state $\boldsymbol{\Omega}^{(\text{new})} = \mathbf{P}_1-\mathbf{x}_1^T(\mathbf{x}_1^T)^\top$.
\end{itemize}

\section{Booked Revenue $\neq$ Measurement}
\label{sec:whynot}
\begin{figure}[H]
  \centering
  \includegraphics[width=0.98\linewidth]{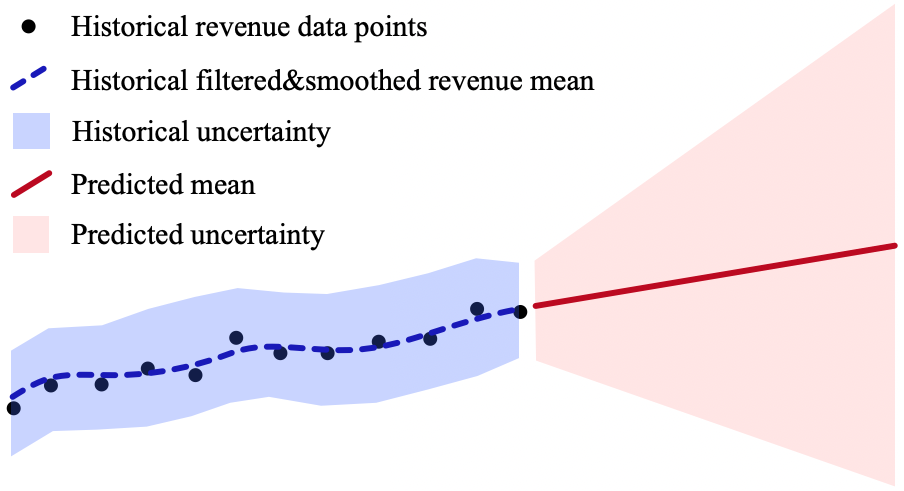}
  \caption{A conceptual explanation describing why we do not directly use booked revenue as measurements.}
  \Description{Need to create one later, this one has copyright problem.}
\label{fig:why_not_book_revenue}
\end{figure}

Obtaining measurements of the revenue signal is crucial to 
model optimization.
One might argue to simply use the data from financial bookings as the measurement of revenue, but this type of revenue measurement is not available in the future.
As illustrated conceptually in Figure~\ref{fig:why_not_book_revenue}, the revenue measurements $y_t$ (black dots) enable optimizing the model parameters and finding the ``true'' revenue signal $x_t$ (blue dashed line).
Although Equation~\eqref{eq:lds_dynamics} allows extrapolating revenue values infinitely into the future, it has two problems: (1) the forecast is degraded to a continuation of the recent historical trend represented as a red line in Figure~\ref{fig:why_not_book_revenue}, and (2) the forecast uncertainty (the red area in Figure~\ref{fig:why_not_book_revenue}) may grow too quickly, making the long-term prediction almost useless. 
As a result, it calls for an approach that can measure the revenue in both past and future time horizons.

\section{Pseudo Code}
\label{sec:pseudo-code}
\subsection{Extrapolate One Revenue Trajectory}
Algorithm~\ref{algo:inner_loop} is the pseudo code to output a $T'$-dimensional vector $[x_{T_{c'}+1}, x_{T_{c'}+2}, \ldots, x_{T_{c'}+T'}]$, formalized in Equation~\eqref{eq:final_revenue_forecasts}, representing one possible revenue trajectory.
The computational complexity of Algorithm~\ref{algo:inner_loop} is approximately $\mathcal{O}(C\times T_c^2)$. 

\subsection{Extrapolate Revenue with Confidence}
Algorithm~\ref{algo:outer_loop} describes the logic to produce the final resulting vector $\hat{\mathfrak{x}}^{(c')}$ formalized in Equation~\eqref{eq:revenue_prediction_company_c_prime}. To generate the confidence estimates, Algorithm~\ref{algo:outer_loop} runs Algorithm~\ref{algo:inner_loop} for $M$ times, followed by an $\mathcal{O}(T')$ confidence estimation step. Since $M$ is a constant and $\mathcal{O}(T')$ has a lower order than $\mathcal{O}(C\times T_c^2)$, the overall computational complexity of SiRE is still $\mathcal{O}(C\times T_c^2)$.

\begin{algorithm}[t]
\footnotesize
\DontPrintSemicolon
\KwInput{Dataset $\mathbf{U}$ formally defined in \eqref{eq:dataset_U} containing revenue time-series of $C$ scaleup companies, a scaleup (noted as $c'$) with a historical time-series
$\{(u_t^{(c')}, b_t^{(c')}, z_t^{(c')}, z_{t+1}^{(c')})|t\!\in\!\mathbb{Z}\!\cap\![1,T_{c'}\!-\!1]\}$}
\KwOutput{$T'$ revenue forecasts for scaleup $c'$: $x_{T_{c'}+1}, x_{T_{c'}+2}, \ldots, x_{T_{c'}+T'}$}
\KwParameter{$\mathbf{A}$ and $\mathbf{c}$ defined in \eqref{eq:matrix_A_c}, number of quantiles $n$
}

\For{$(t' = 1; t' \leq T_{c'}; t'++)$}{

    Assemble a set ${\mathbf{U}_{t'}^{(c')}}$ from $\mathbf{U}$ using Equation \eqref{eq:dataset_measuring_subset};
    
    Collect a revenue growth set $\grave{\mathbf{Z}}$ from $\mathbf{U}_{t'}^{(c')}$ using Equation \eqref{eq:Z_in_measuring_subset};
    
    Divide $\grave{\mathbf{Z}}$ into $n$ quantiles with $n+1$ boundaries $q_0, q_1, \ldots, q_{n-1}, q_n$;
    
    Get the measuring dataset $\overline{\mathbf{U}}$ following Equation \eqref{eq:dataset_growth_subset};
    
    Extract all probable next revenue growth to form a set $\acute{\mathbf{Z}}$: Equation \eqref{eq:Z_next_in_U_bar};
    
    Sample a revenue growth $\hat{z}$ using Equations \eqref{eq:sample_next_growth};
    
    Obtain the measured revenue $y_{t'+1}^{(c')}$ using Equation \eqref{eq:final_measurement};
    
}

Initialize $\mathbf{Q}$, $\mathbf{R}$, $\boldsymbol{\Omega}$, and $\boldsymbol{\mu}$ with \eqref{eq:init_Q_R_Omega} and \eqref{eq:init_mu};

\For{$(i = 1; i \leq 10; i++)$} {

    \For{$(t' = 1; t' \leq T_{c'}; t'++)$}{
        Forward filtering following \eqref{eq:forward_filtering};
    }
    
    \For{$(t' = T_{c'}; t' \geq 1; t'--)$}{
        Backward smoothing following \eqref{eq:backward_smoothing_1} and \eqref{eq:backward_smoothing_2};
    }
    
    Calculate new values for $\mathbf{R}$, $\mathbf{Q}$, $\boldsymbol{\mu}$, and $\boldsymbol{\Omega}$ following Appendix~\ref{sec:m-step};
  
}

Calculate $\mathbf{x}_{T_{c'}+1}^{(c')}$ via forward filtering expressed in \eqref{eq:forward_filtering} w.r.t. $t=T_{c'}+1$;

\For{$(\tau = T_{c'}+1; \tau \leq T_{c'}+T'-1; \tau++)$} {

    Estimate revenue growth $z_{\tau}^{(c')}$ with \eqref{eq:revenue_growth_in_prediction}; 
    
    Assemble a set $\mathbf{U}_{\tau}^{(c')}$ from $\mathbf{U}$ using Equation \eqref{eq:forecast_measuring_dataset};
    
    Collect a revenue growth set $\grave{\mathbf{Z}}$ from $\mathbf{U}_{\tau}^{(c')}$: similar to Equation \eqref{eq:Z_in_measuring_subset};
    
    Divide $\grave{\mathbf{Z}}$ into $n$ quantiles with $n+1$ boundaries $q_0, q_1, \ldots, q_{n-1}, q_n$;
    
    Get a measuring dataset $\overline{\mathbf{U}}$ following Equation \eqref{eq:forecast_growth_subset};
    
    Sample a revenue growth $\hat{z}$ following Equations \eqref{eq:Z_next_in_U_bar} to \eqref{eq:sample_next_growth};
    
    Calculate measured revenue $y_{\tau+1}^{(c')}$ using Equation \eqref{eq:forecast_final_measurement};
    
    Calculate $\mathbf{x}_{\tau+1}^{\tau+1}$, $\mathbf{P}_{\tau+1}^{\tau}$ and $\mathbf{P}_{\tau+1}^{\tau+1}$ with a filtering operation depicted in \eqref{eq:forecast_filtering};
  
}

\For{$(\tilde{t} = T_{c'}+T'; \tilde{t} \geq 1; \tilde{t}--)$} {

    Calculate $\mathbf{x}_{\tilde{t}-1}^{T_{c'}+T'}$ with the smoothing procedure in \eqref{eq:global_smoothing_in_forecasting};
}

Extract revenue forecasts $x_{T_{c'}+1}, x_{T_{c'}+2}, \ldots, x_{T_{c'}+T'}$ following \eqref{eq:final_revenue_forecasts};

\textbf{return} $[x_{T_{c'}+1}, x_{T_{c'}+2}, \ldots, x_{T_{c'}+T'}]$

\caption{Forecast $T'$ revenues for scaleup company $c'$}
\label{algo:inner_loop}
\end{algorithm}

\begin{algorithm}[t]
\footnotesize
\DontPrintSemicolon
\KwInput{Dataset $\mathbf{U}$ formally defined in \eqref{eq:dataset_U} containing revenue time-series of $C$ scaleup companies, a scaleup (noted as $c'$) with a historical time-series
$\{(u_t^{(c')}, b_t^{(c')}, z_t^{(c')}, z_{t+1}^{(c')})|t\!\in\!\mathbb{Z}\!\cap\![1,T_{c'}\!-\!1]\}$}
\KwOutput{Revenue prediction $\hat{\mathfrak{x}}^{(c')}$ formalized in \eqref{eq:revenue_prediction_company_c_prime}: the $T'$ revenue forecasts with 95\% confidence band}

\KwParameter{$\mathbf{A}$ and $\mathbf{c}$ defined in \eqref{eq:matrix_A_c}, the number of quantiles $n$, and the number of prediction trails $M$}
    
Initialize $\widehat{\mathbf{X}}^{(c')}=[]$;

\For{$(m = 1; m \leq M; m++)$}{
    
    Calculate $\widehat{\mathbf{X}}^{(c')}_{\langle m, \cdot \rangle}=\left[x_{T_{c'}+1}^{\langle m \rangle}, x_{T_{c'}+2}^{\langle m \rangle}, \ldots, x_{T_{c'}+T'}^{\langle m \rangle}\right]$ with Algorithm~\ref{algo:inner_loop};
    
    Append $\widehat{\mathbf{X}}^{(c')}_{\langle m, \cdot \rangle}$ as a new row in $\widehat{\mathbf{X}}^{(c')}$;
}

Initialize $\hat{\mathfrak{x}}^{(c')}=[]$;

\For{$(t' = 1; t' \leq T'; t'++)$}{

    Extract the $t'$-th column from $\widehat{\mathbf{X}}^{(c')}$, i.e. $\widehat{\mathbf{X}}^{(c')}_{\langle \cdot, t' \rangle}$=$[x_{T_{c'}+t'}^{\langle 1 \rangle}, \ldots, x_{T_{c'}+t'}^{\langle M \rangle}]^{\top}$;
    
    Compute mean revenue $\bar{x}_{T_{c'}+t'}$ by averaging values in $\widehat{\mathbf{X}}^{(c')}_{\langle \cdot, t' \rangle}$;
    
    Compute error margin $\beta_{T_{c'}+t'}$ from $\widehat{\mathbf{X}}^{(c')}_{\langle \cdot, t' \rangle}$ following Equation \eqref{eq:confidence_interval_beta};
    
    Append $\bar{x}_{T_{c'}+t'}\pm\beta_{T_{c'}+t'}$ to $\hat{\mathfrak{x}}^{(c')}$;
}

\textbf{return} $\hat{\mathfrak{x}}^{(c')}$ in the form of Equation~\eqref{eq:revenue_prediction_company_c_prime};

\caption{Forecast $T'$ revenues with confidence estimate for a scaleup company $c'$}
\label{algo:outer_loop}
\end{algorithm}

\end{document}